\documentclass[floatfix,prd,twocolumn,letterpaper,lengthcheck,superscriptaddress,showpacs,amssymb,amsmath,amsfonts,aps,altaffilletter,nofootinbib,nopreprintnumbers,showpacs,longbibliography]{revtex4-1}
\usepackage[english]{babel}
\usepackage[utf8]{inputenc}
\usepackage[colorinlistoftodos, color=green!40, prependcaption]{todonotes}
\usepackage{xcolor}
\usepackage{adjustbox}
\usepackage[T1]{fontenc}
\usepackage{lipsum}
\usepackage{csquotes}
\usepackage{hyperref}
\usepackage{amsmath}
\usepackage{graphicx}
\usepackage{caption}
\usepackage{subcaption}
\usepackage[colorinlistoftodos]{todonotes}
\usepackage{float}
\usepackage{mathtools}
\usepackage{bigints}

\begin{document}
\title{Ecosystem for Closed Timelike Curves: An Energy Conditions Perspective}
\author{Sashideep Gutti}
\email{sashideep@hyderabad.bits-pilani.ac.in}
\affiliation{Birla Institute of Technology and Science, Pilani—Hyderabad Campus, Hyderabad, India 500 078 }

\author{Shailesh Kulkarni}
\email{shailesh@physics.unipune.ac.in}
\affiliation{${}^b$Department of Physics, Savitribai Phule Pune University, Ganeshkhind, Pune 411 007, India}

\author{Vaishak Prasad}
\email{vaishak@iucaa.in} 
\affiliation{${}^c$ Inter-University Centre for Astronomy and Astrophysics, Post Bag 4, Ganeshkhind, Pune 411 007, India}

\author{Sheldon Suresh}
\email{sureshsheldon@gmail.com}
\affiliation{Saint Xaviers, Fort, Mumbai, Maharashtra 400001, India}


\begin{abstract}
In this article, we explore the relationship between the existence of closed timelike curves and energy conditions that occur in the Kerr-Newman spacetime. To quantify the dependence, we define a correlation index between energy conditions and closed timelike curves. Based on the inputs from Hawking's chronology protection conjecture, we analyze two popular variants of Kerr-Newman spacetime: Non-commutative and Rastall Kerr-Newman spacetimes. These two models provide complementary scenarios that aid in analyzing Hawking's statements regarding the correlation of closed timelike curves and energy conditions from a local and a global perspective. We report the results outlining the possible role played by violations of energy conditions in eliminating the closed timelike curves in two contrasting situations, namely in spacetimes with and without curvature singularities. 
\end{abstract}

\maketitle

\section{Introduction}

The formulation of the basic laws of physics intrinsically assumes the preservation of causality (the causal ordering of events).  A Closed Timelike Curve (CTC) is a closed curve whose tangent is everywhere timelike with topology $S^1$. The existence of CTCs in a spacetime, therefore, implies the violation of cause/effect relations that are sacrosanct to predictability in physics \cite{Morris:1988tu}. To deal with the pathology,  Hawking came up with the Chronology Protection Conjecture which states that {\it The laws of physics do not allow the appearance of closed timelike curves}. From a classical general relativity perspective, many spacetimes have the pathology of the CTCs; these are bonafide solutions to the Einstein's field equations e.g G\"{o}tt, G\"{o}del, Kerr family of spacetimes 
\cite{Gott:1998re,Gott:1990zr, godel1} (see also \cite{Hawking:1973uf} and references therein). More prominent are the CTCs in the Kerr-Newman spacetimes. The CTCs are usually present in the interior of the inner horizon in the Kerr-Newman spacetime (they are also present in the Kerr-Newman spacetime that has a globally naked singularity).  The CTCs in this spacetime are in causal contact with the singularity and therefore the buck of unpredictability rests with the singularity. The quotation by Hawking reiterates the fact. Quoting Hawking from his famous paper \cite{Hawking:1991nk}, "In spacetimes exhibiting closed timelike curves, either there exist singularities (like in the Kerr spacetime), pathological behaviour at infinity (as in the G\"{o}del and G\"{o}tt spacetimes) or the weak energy condition is violated". 

The singularity of spacetime is a well known and more dangerous pathology that the theory of relativity is plagued with. There is an extensive study on reconciling the singularities with predictability in physics. The cosmic censorship conjecture by Penrose \cite{penrose} states that the singularities are always hidden behind a horizon, therefore, protecting large portions of spacetime from the unpredictability that arises due to causal communication with the singularity. Though there is no formal proof yet regarding the conjecture, there are many models proposed with the intention of proving/disproving the same. A review of the status of the problem can be found in  \cite{krolak1, krolak2, joshi1, joshi2, tpsingh, landsman}. The models that present the possibilities of naked singularities (where the causal rays from the singularity can reach non-singular points in spacetime) are not generic enough to conclude against the conjecture. It is well accepted that once we have a quantum theory of gravity, the singularities are expected to be smeared away into a high curvature region suggesting that the singularities are the artefacts of classical general relativity \cite{thiemann,bojowald,modestosingularity,hussain}. 

Although  substantial efforts have been made (e.g. see \cite{Visser:1995cc,Friedman:2006tea})
to understand the issue related to  CTCs, it still lacks the rigorous status as the singularity theorems associated with curvature singularities. The existence, formation and avoidance of CTCs in spacetime are poorly understood. One factor that is detrimental to exhaustive research in the area of CTCs is the lack of an adequate number of models that present the pathology of CTCs. Even though a lot of work is carried out for regular black hole geometries where there are models with the singularities are smeared out, the elimination of CTCs are usually not addressed. Our recent work \cite{Prasad:2018hdj} is the first attempt to address the issue of eliminating CTCs using variants of gravity.  The ecosystem for the existence/emergence of the CTCs in spacetime is not well understood and the field of research is in its nascent stage. Since quantum gravity is expected to smear out the singularities. It is maybe reasonable to expect that the solutions within the correct quantized theory of gravity theory will also be free of the CTCs. Some peculiarities in the spacetimes endowed with CTC that appears at the semi-classical level have been studied extensively in  
\cite{Boulware:1992pm, Visser:2002ua}.
We, in this article, address this elephant in the Kerr-Newman space-time and track its ecosystem from the point of view of energy conditions.

In the Kerr spacetime, the CTCs exist in an environment where the energy conditions are preserved, but in the presence of a curvature singularity. However, various variants of the Kerr family of solutions have been found in modified gravity models that are interesting for their own reasons. In \cite{Prasad:2018hdj}, we have analyzed the non-commutative version of the Kerr-Newman spacetime presented in \cite{Modesto:2010rv}. It is shown that there exists a parameter space that constitutes the Kerr-Newman spacetime, ($M,Q,a, \Theta$) for which the pathology due to the presence of CTC exists and a parameter space in which the CTCs are eliminated. We showed that for a given set of parameters $(M,Q,a)$, we can choose the non-commutative parameter $\Theta=\Theta_0$ such that CTCs are eliminated. The paper also describes the analysis of the Kerr-Newman version in $f(R)$ gravity and discusses the possibility of elimination of the CTCs within the parameters present in the model. 
A systematic study of CTCs, their existence and elimination is, therefore, a gap in current research. It is well known that the energy conditions are violated in some quantum field theoretic contexts (e.g Casimir energy). It is also fairly certain that for eliminating the spacetime singularities, quantum gravity is instrumental. Energy condition violation is therefore expected to prevent the formation of singularities. A lot of work on the regular black holes and energy conditions have been carried out in \cite{nves,bambi, tom}  with proposals for singularity free spacetimes. Most of these models are based on insights from modified general relativity.

One can ask the following questions about the CTCs and their elimination. Does avoidance of singularities automatically lead to the elimination of CTCs as well? Are energy condition violations needed if the final aim is to eliminate CTCs that are present in the solutions for classical general relativity? Can we preserve energy conditions and eliminate CTCs? Among the different energy conditions  (weak, strong, null, dominant, etc), which of them is crucial for the preservation and elimination of CTCs?

Variants of Kerr-Newman spacetimes present us with an excellent laboratory to
test out the dependence of CTCs on the various energy conditions and provide answers to the above questions.
It can be expected that the prevention or elimination of CTCs might be possible provided we compromise one (or more) energy conditions.
 
In this work, we study the correlation between energy conditions and CTCs. We study this correlation in two ways, local and global perspectives. Hawking's statement connects the violation of the energy conditions and the presence of CTCs. We explore the possibility of refining this connection in the following way. Suppose we consider a point in spacetime through which a CTC passes. 
Some of the energy conditions may not be violated locally in the region of CTCs (i.e. at the location of CTCs), but rather globally. It is therefore worthwhile  
to investigate the interplay between the energy conditions and CTCs from the local as well as global points of view. 
The understanding of this correlation can be very useful in many contexts. Firstly, when one considers models for modified gravity, the selection of one model over the other is usually based on consistency with cosmological observations. One more parameter for this selection criteria could be the CTC elimination. If one constructs equivalent models of certain solutions, e.g the Kerr-Newman spacetime, using phenomenological quantum gravity models, the status of causality preservation due to elimination of CTCs could be yet another independent criterion for choosing one model over the other.
 
We study two cases of modified Kerr-Newman spacetime: The non-commutative (regular) Kerr-Newman spacetime and the Kerr-Newman version found in Rastall gravity. Hawking's statement brings out two distinct situations. CTCs near singularity and CTCs in the absence of singularities. The two models presented here cover both these scenarios. The Rastall version has curvature singularity and therefore CTCs exist in the presence of a curvature singularity. We show that one can choose parameters in this model to eliminate CTCs. The non-commutative counterpart is regular and the curvature singularity is smeared out. So this presents a model that has CTCs but no curvature singularity. Both these models, therefore, present excellent theoretical testing grounds to examine the ecosystem of CTCs. 

The paper is organized as follows. In Sec.~\ref{sec:cindex}, we propose a general formalism for quantifying the association of energy conditions with the existence of closed timelike curves. In Secs.~\ref{sec:nckn} and \ref{sec:rastall} we apply this procedure to the variants of Kerr-Newman spacetimes in non-commutative and Rastall gravity and discuss the correlations between the energy conditions and CTCs. In Sec.~\ref{sec:con} we make concluding remarks.

Throughout, we use geometric units in which $G=c=1$. We also set the ADM mass of the spacetime to be 1 i.e. $M=1$. For algebraic computations, we use \emph{Mathematica} \cite{math}. Although we deal with solutions that are used to describe the exterior geometries of black holes, we will refer to them as spacetimes as for some of the parameters, the spacetime may or may not contain horizons.

\section{Correlation index for the CTCS}\label{sec:cindex}
We explore in the subsequent sections, the correlation of the CTCs with the energy conditions. Our goal is to explore the models in variants of general relativity. To make the estimate quantitative,  we give the prescription for defining a correlation index in this section. We observe a common feature among the variants of Kerr-Newman spacetime solutions. The variants are in general, dependent on few parameters $(\mu_1, \mu_2...\mu_n)$ that are specific to the model of gravity chosen. The Einstein gravity corresponds to particular values of these parameters. The Kerr-Newman metric has an axial symmetry corresponding to a Killing vector $\xi_{(\phi)}^{\alpha}$ presented in a coordinate chart say $(x^{\alpha})$ (Here the subscript $\phi$ is a label indicating the axial Killing vector field). The CTCs in these coordinates can be singled out by examining the sign of the norm of this Killing vector $\xi_{(\phi)}^{\alpha}\xi_{(\phi) \alpha}$. Wherever  $\xi_{(\phi)}^{\alpha}\xi_{(\phi) \alpha}$ becomes negative, we have CTCs. Now this expression for the norm is a scalar function of the coordinates ($x^{\alpha}$). The modified Kerr-Newman solution is generally dependent on the various parameters $(\mu_1, \mu_2...\mu_n)$ and  so is the norm of $\xi_{(\phi)})^{\alpha}$.  When one computes the expression for the energy conditions, one gets a function of the form $f(x^{\alpha}, \mu_1,\mu_2...\mu_n)$. The sign of that function dictates  whether a particular energy condition is violated or not.  Motivated by this we {\it associate} the index for correlation between the existence of CTCs and energy conditions as
\begin{equation}
\label{cindex1}
    \mathcal{C}=\frac{\bigint \frac{f}{|f|}\left(\frac{\xi_{(\phi)}^{\alpha}\xi_{(\phi) \alpha}}{|\xi_{(\phi)}^{\alpha}\xi_{(\phi) \alpha}|}-1\right) \sqrt{h} \ dX \ d\mu
    }{\bigint \left(\frac{\xi_{(\phi)}^{\alpha} {\xi_{(\phi)}}_{\alpha}}{|\xi_{(\phi)}^{\alpha}\xi_{(\phi) \alpha}|}-1\right) \sqrt{h} \ dX \  d\mu} 
\end{equation}
Here $dX$ is a short notation to the relevant spacetime volume element,  $d\mu$ stands for volume in the parameter space $d\mu_1d\mu_2 \cdots d\mu_n$ and $\sqrt{h}$ is the infinitesimal volume element on the relevant hypersurface orthogonal to a timelike Killing vector field $\xi_{(t)}$. If we use the standard Boyer-Lindquist coordinate system ($t,r, \theta, \phi$), the norm $\xi^{\alpha}_{(\phi)}\xi_{(\phi )\alpha}$ is then equal to the metric component $g_{\phi\phi}$. In the models that we consider below, we restrict our analysis to the equatorial plane and hence we set $\theta=\pi/2$. Due to axial symmetry we can further set $\phi=0$. Consequently, the index now assumes the form,
\begin{equation}
\label{cindex2}
    \mathcal{C}=\frac{\bigint \frac{f}{| f|}\left(\frac{g_{\phi\phi}}{|g_{\phi\phi}|}-1\right) \sqrt{h} \ dX \ d\mu
    }{\bigint \left(\frac{g_{\phi\phi}}{|g_{\phi\phi}|}-1\right) \sqrt{h} \ dX \ d\mu} 
\end{equation}
The motivation for the above definition of the index is stated below. The term $\displaystyle \left(\frac{g_{\phi\phi}}{|g_{\phi\phi}|}-1\right)$ is zero wherever CTCs are not present and is equal to $-2$ in regions where CTCs are present. The term $f/|f|$ is $+1$ whenever the relevant energy condition is preserved, as we vary the spacetime coordinates as well as the parameters $\mu_i$. The term  $f/|f|$ will then have a value $-1$ wherever the relevant energy condition is violated. Hence, the numerator of the integrand in Eq.~(\ref{cindex2}) is positive for the parameters for which the CTCs are present and relevant energy condition is violated. It is zero wherever CTCs are absent. It is $-2$ wherever CTCs are present and the relevant energy conditions are preserved.
The denominator of the above expression integrates over the entire parameter range which includes CTCs and it gives vanishing contribution from the points where  CTCs are absent. Consequently, the index is ill-defined in situations when there are no CTCs exist anywhere in the spacetime for all parameter values. The resultant index, therefore, takes values between $-1$ and $+1$. If the index is negative, one can infer that CTCs prefer to exist in the region where the relevant energy condition, denoted by $f(x^{\alpha}, \mu_1,\mu_2...\mu_n)$ is violated. A positive value of the index implies that CTCs prefer to live in an energy condition preserved environment suggesting that elimination of CTCs might require violation of energy conditions. \\
In the models below, we evaluate the index given by Eq.~\eqref{cindex2} in a sub-region of the parameter space, to systematically study the correlation between the existence of CTCs and the weak energy condition.
To avoid numerical artefacts, while implementing the index formula, care must be taken such that CTCs are present in the domain of integration.
\section{Non-commutative Kerr-Newman spacetime}\label{sec:nckn}
We consider charged rotating black hole spacetimes
inspired by non-commutative geometry. 
The motivation behind introducing the 
non-commutativity in the usual black holes
setup is to cure pointlike curvature
singularities (e.g. the Dirac-delta like the singularity of
Schwarzschild BH, the ring singularity of the Kerr BH) by
effectively replacing it by appropriate mass distribution \cite{Nicolini:2005vd, Nicolini:2009gw, Spallucci:2008ez}. Within the framework of
coordinates coherent state approach together with expected first
order quantum gravitational correction to classical general
relativity it can be shown that the Dirac-delta like singularities
will be replaced by Gaussian distributions \cite{Modesto:2010rv}. For the point particle at
the origin, the non-commutative nature of coordinates reflects into a modification to usual 
delta functions distribution $\delta({\bf x})$ as

\begin{equation}
    \rho_{0}({\bf x}) = \frac{1}{2\pi \Theta}e^{-\frac{{\bf x}^2}{2\Theta}} 
\end{equation}
The width of the Gaussian distribution is now characterized
by the Non-Nommutative (NC) parameter $\Theta$.  
It has  been shown that \cite{Cembranos:2011sr} the effective
corrections to the Einstein’s-field equations due to the 
above replacement can be modelled by replacing the
pointlike sources by a suitable Gaussian distribution while
keeping the differential operators 
unchanged. The non-commutative solution for the Kerr-Newman (KN) spacetime   
can be written as \cite{Modesto:2010rv}
\begin{eqnarray}
ds^2 &=& - \frac{\Delta - a^2 \sin^2 \theta}{\rho^2}dt^2 - 2a \sin^2 \theta\Big(
1 - \frac{\Delta - a^2 \sin^2 \theta}{\rho^2}\Big)dt d\phi  \nonumber \\ 
&& + \frac{\rho^2}{\Delta}dr^2 + \rho^2 d\theta^2 + 
\frac{\Sigma^2}{\rho^2}\sin^{2}\theta d\phi^2
\end{eqnarray}
with 
\begin{eqnarray}
\Delta &=& r^2 - 2 m(r) r + a^2 + q^{2}(r) \nonumber\\
\Sigma^2 &=& (r^2 + a^2)^2 - a^2 \Delta \sin^{2}\theta \\
\rho^2 &=& r^2 + a^2 \cos^{2}\theta \nonumber
\end{eqnarray}
where the mass and charge functions take the form
\begin{equation}
     m(r) = M \frac{\gamma(3/2, r^2/(4\Theta))}{\Gamma(3/2)}
\end{equation}
\begin{eqnarray}
 q^{2}(r) &=& \frac{Q^2}{\pi}\Big[\gamma^2(1/2, r^2/(4\Theta)) 
-\frac{r}{\sqrt{2\Theta}}\gamma(1/2, r^2/(2\Theta)) \nonumber \\
&+& r \sqrt{\frac{2}{\Theta}}
\gamma(3/2, r^2/(4\Theta))\Big]
\end{eqnarray}
Using the Einstein equations for the above metric the expression for components of energy-momentum tensor read as  
\begin{align}
T^{\mu}_{\nu} = \begin{bmatrix}
T^{t}_{t} & 0 & 0 & T^{t}_{\phi}\\
0 & T^{r}_{r} & 0 & 0\\
0 & 0 & T^{\theta}_{\theta} & 0\\
T^{\phi}_{t} & 0 & 0 & T^{\phi}_{\phi}
\end{bmatrix}
\label{EM1}
\end{align}
At this point, we would like to make the following observations. 
The energy-momentum tensor is off-diagonal in the $t-\phi$ sector. To extract the information about the energy conditions we shall seek a tetrad frame in which $t-\phi$ sector is diagonalised while keeping the other components intact since other components are already in diagonal form. It is worth mentioning that our energy-momentum tensor Eq.~\eqref{EM1} falls under the type-I of Hawking-Ellis classification
\cite{Hawking:1973uf} which is based on the extent to which the orthonormal components of the energy-momentum tensor can be expressed in a diagonal form \cite{Martin-Moruno:2017exc}. 
General expressions for the components of energy momentum tensor are complicated and are given in the appendix. Since the region near to the  equator ($\theta =\pi/2$) is densely populated with CTCs, we shall restrict ourselves in the near equatorial region\footnote{This point is explained with plots at the beginning of Sec.~\ref{sec:sec4}}. For $\theta = \pi/2$, the energy momentum tensor can be written as 

\begin{widetext}
\begin{eqnarray}
8\pi T^{t}_{t} &=& -\frac{1}{4 r^6}\Big[
-4 a^2 r^3 m''(r,\Theta )+\Big(8 a^2 r^2+8 r^4\Big) m'(r,\Theta )+\nonumber\\
   && 2 a^2 r^2 q''(r,\Theta ,Q)+   2 r \Big(-4 a^2-2 r^2\Big)
   q'(r,\Theta ,Q)+\Big(8 a^2+4 r^2\Big) q(r,\Theta
   ,Q)\Big]\nonumber\\
   8\pi T^{r}_{r}&=& -\frac{r \left[2 r m'(r,\Theta )-q'(r,\Theta
   ,Q)\right]+q(r,\Theta ,Q)}{r^4} \nonumber\\
   8\pi T^{\theta}_{\theta} &=& \frac{1}{4r^4}\Big[2 r^2 \left(-4 m'(r,\Theta )-2 r m''(r,\Theta
   )+q''(r,\Theta ,Q)\right) \nonumber\\ 
   && + 4 \left(r \left(2 r
   m'(r,\Theta ) -q'(r,\Theta ,Q)\right)+q(r,\Theta
   ,Q)\right)\Big] \nonumber\\
8\pi T^{\phi}_{\phi} &=& -\frac{1}{4 r^6}\Big[2 r \left(2 r^2 \left(a^2+r^2\right) m''(r,\Theta
   )+\left(4 a^2+2 r^2\right) q'(r,\Theta ,Q)\right)-
   \nonumber\\ 
   && 8a^2 r^2 m'(r,\Theta )-2 r^2 \left(a^2+r^2\right)
   q''(r,\Theta ,Q)+2 \left(-4 a^2-2 r^2\right)
   q(r,\Theta ,Q)\Big] \nonumber \\  
  8\pi T^{t}_{\phi} &=& \frac{1}{4r^6}\Big[a \left(a^2+r^2\right) \Big(-4 r^3 m''(r,\Theta )+8 r^2 m'(r,\Theta )+2 r^2 q''(r,\Theta ,Q)\nonumber\\ 
    && -8 r q'(r,\Theta ,Q)+8 q(r,\Theta ,Q)\Big)\Big] \nonumber \\  
   8\pi T^{\phi}_{t} &=& \frac{1}{4r^6}
   \Big[2 a r^2 \left(4 m'(r,\Theta )+2 r m''(r,\Theta
   )-q''(r,\Theta ,Q)\right)\nonumber\\
    &&-8 a \left(r \left(2 r m'(r,\Theta )-q'(r,\Theta ,Q)\right)+q(r,\Theta,Q)\right)\Big]\nonumber
\end{eqnarray}
\end{widetext}

 Next, we carry out the reduction of the energy momentum tensor to the canonical form. This requires us to find the eigenvalues of the expression 
\begin{equation}
det|T_{\mu\nu}-\lambda g_{\mu\nu}|=0
\label{det}
\end{equation}
Using the above one can reduce the energy-momentum tensor to the canonical form. If the eigenvalues are real, then the energy-momentum tensor belongs to type-I according to Hawking and Ellis classification. type-I implies that the energy-momentum tensor and the metric are simultaneously diagonalizable. This implies that one can find a tetrad basis in which the canonical energy-momentum tensor is 
$T_{ab}=T_{\mu\nu}e^{\mu}_a e^{\nu}_b$ where ($e^{\mu}_a$) are the tetrads which are the simultaneous eigenvectors of type-I energy-momentum tensor and the metric. Thus,
the eigenvalues represent the principle pressures and density. 
For the NCKN case, we show that the energy-momentum tensor can be canonically decomposed to type-I by Hawking classification on the equatorial plane $\theta=\pi/2$. We evaluate the eigenvalues and eigenvectors ($e_{(t)}$, $e_{(\phi)}$) by focussing on the $t-\phi$ sector of the energy-momentum tensor (the rest of the components are already in the canonical form). The eigenvector $e_{(t)}$ is timelike while $e_{(\phi)}$ is spacelike. The explicit form of these tetrads is furnished by the following expressions 

\begin{align}
    e^{t}_{(t)} &=  \frac{a^2 + r^2}{a} \ ; \ e^{\phi}_{(t)} = 1 \ ; \\ 
    e^{t}_{(\phi)} &= a \ ; \ e^{\phi}_{(\phi)} = 1  \ ; \  e^{r}_{(t)} = \ ; \  e^{\theta}_{(t)} =  \ ; \  e^{r}_{(\phi)} = \ ; \  e^{\theta}_{(\phi)} = 0 
\end{align}
Also, as expected, the eigenvectors are orthogonal to each other. After normalizing the eigenvectors and re-expressing the components of the energy-momentum tensor in the $t-\phi$ sector, we finally obtain the following components 
\begin{eqnarray}
8\pi \tilde{T}^{t}_{t} &=& -\frac{1}{r^4}\Big[q(r,\Theta,Q)+2r^2 m'(r,\Theta) -rq'(r,\Theta,Q)\Big]\nonumber\\
8\pi \tilde{T}^{\phi}_{\phi} &=& \frac{1}{r^4}\Big[q(r,\Theta,Q)-r^3 m''(r,\Theta) -r q'(r,\Theta,Q)\nonumber\\
&+&2r^2 q''(r,\Theta,Q)\Big]\nonumber\\
8\pi \tilde{T}^{t}_{\phi} &=& 8\pi \tilde{T}^{\phi}_{t} = 0 
\end{eqnarray}
while other components remain unchanged.
\subsection{Correlation of energy conditions with presence of CTCs} \label{sec:sec4}
 As demonstrated in \cite{Prasad:2018hdj}, for a given charge, angular momentum and mass of a black hole, we can choose the non-commutative parameter ($\Theta>\Theta_{critical}$) such that the CTCs are eliminated. In contrast, the usual Kerr-Newman spacetime has CTCs in its spacetime for the same set of parameters (this can be achieved by taking the limit $\Theta=0$). 
 
\begin{figure}
  \includegraphics[width=\columnwidth]{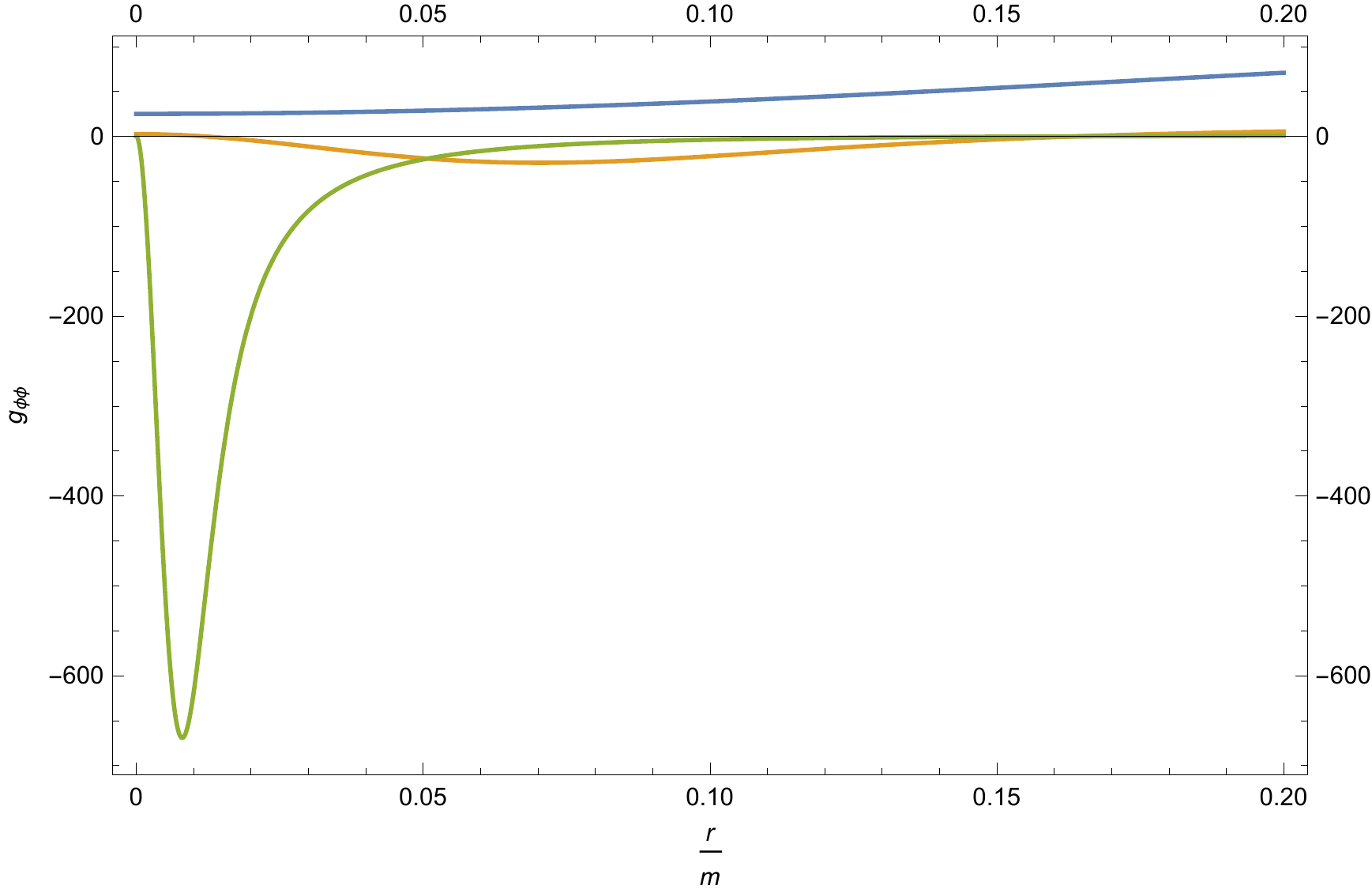}
  \caption{Behaviour of $g_{\phi\phi}$ in the equatorial plane ($\theta =\pi/2$)  for the fixed value of charge $Q= 0.6$ and $a=0.5$ with $\Theta=10^{-2}$ (blue), $\Theta=10^{-3}$ (orange) and $\Theta=10^{-5}$ (green). Region (in $r$ space) containing the closed time like curves exists for orange and blue curves.}
  \label{fig:ctc_reg_1}
\end{figure}

\begin{figure}
  \includegraphics[width=0.75 \columnwidth]{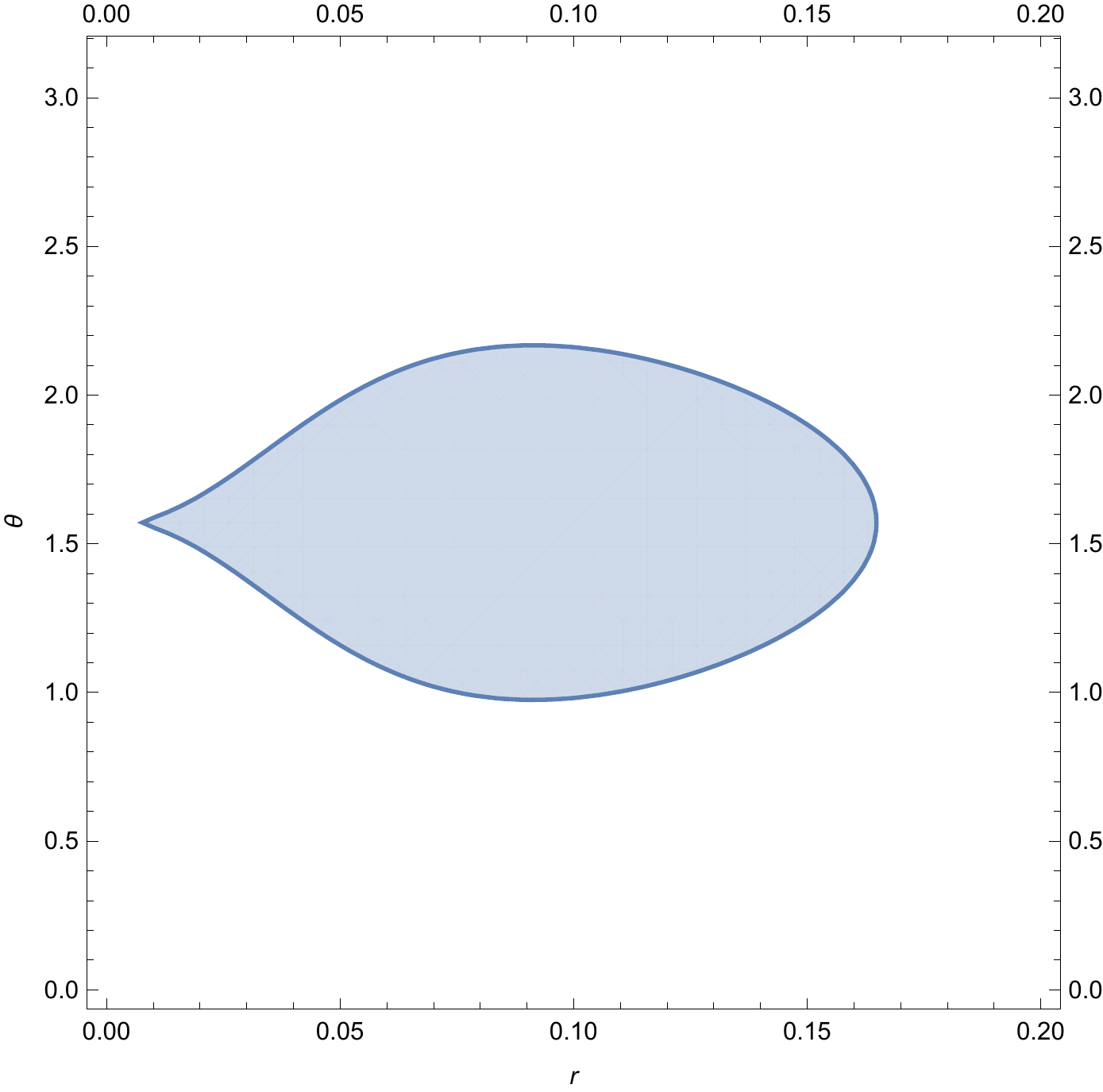}
   \includegraphics[width=0.75 \columnwidth]{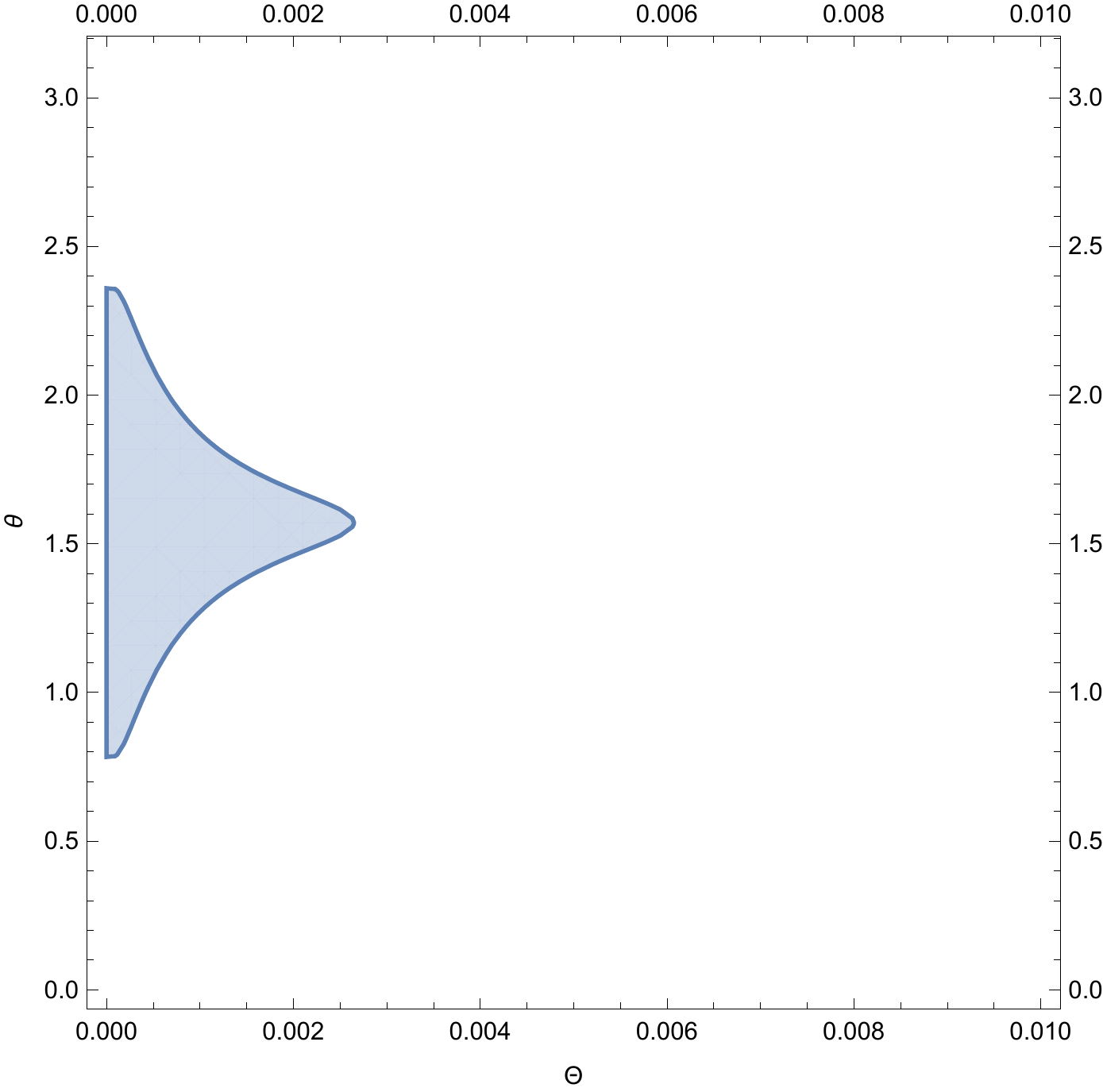}

  \caption{Top: CTC region in $r-\theta$ space. Bottom: CTC region in $\theta-\Theta$ space}
  \label{fig:ctcr}
\end{figure}

 In Fig.~\ref{fig:ctc_reg_1} we have plotted  $g_{\phi\phi}$ against the re-scaled  radial coordinate $r$ for three different values of non-commutative parameter $\Theta$ while keeping $Q$ and $a$ fixed. For $\Theta=10^{-3}$ (orange curve) and $10^{-5}$ (green curve) there are regions of CTC, with different radial extent and depth. We see that CTCs are absent close to $r=0$ and develop for a larger radius. Though not apparent from the figure, the CTCs are again absent for a larger radius $r$ since the metric approaches the usual Kerr-Newman spacetime asymptotically \cite{Prasad:2018hdj,Modesto:2010rv}. We also observe that the effect of non-commutativity is more prominent at small radii and indicates that the non-commutativity is conducive to eliminating the CTCs. 
 In Fig.~\ref{fig:ctcr}, we show that the CTCs are more prominent on the equatorial plane. On the top panel of Fig.~\ref{fig:ctcr}, the $x$-axis is radial coordinate $r$ and $y$-axis is the angle $\theta$ in radians.  The shaded portion is the region where CTCs are present. At the bottom panel, we plot $\Theta$ on $x-$ axis and $\theta$ on $y-$ axis for  fixed radius $r$. This plot illustrates that CTCs are concentrated more near $\theta=\pi/2$. 
For correlating this information with the energy condition, we similarly obtain the violation details of the energy conditions as a function of $r$. We then superimposed on the same plot, the CTC region and the energy condition violation region. The correlation obtained is therefore a local one where we address the question that for a given point, the presence of CTCs is correlated with the violation of a particular energy condition or not.
We present here the observations regarding the possible correlation of the CTCs and energy conditions. For the type I matter field the various energy conditions are given by 
\begin{eqnarray}
&&\rho \geq 0 \ , \rho + p_{i} \geq 0 \qquad \text{WEC} \nonumber\\ 
&& \rho + \sum_{i=1}^{3} p_{i} \geq 0 \qquad \qquad \text{SEC} \nonumber\\
&&\rho + p_{i} \geq 0  \ \qquad \qquad \quad \text{NEC} \nonumber\\
&&\rho - |p_{i}| \geq 0 \qquad \qquad \quad \text{DEC}\nonumber
\end{eqnarray}

\begin{figure}
  \includegraphics[width=\columnwidth]{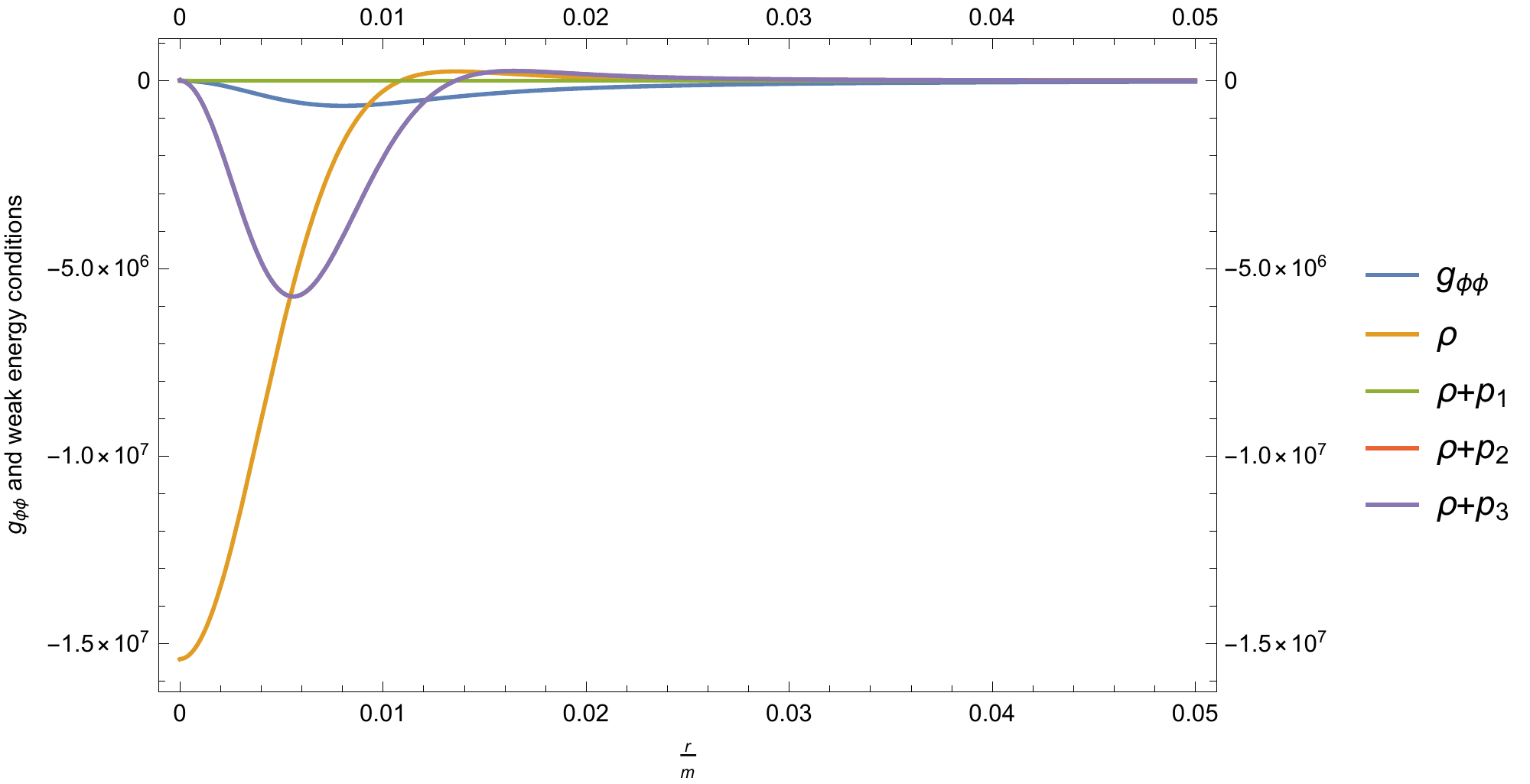}
  \caption{Behaviour of $g_{\phi\phi}$ and weak energy conditions against $r$. The parameters are set to: $\Theta=10^{-5}, Q=0.6$ and $a=0.5$}
  \label{fig:ctcweak}
\end{figure}

\begin{figure}
\includegraphics[width=0.75\columnwidth]{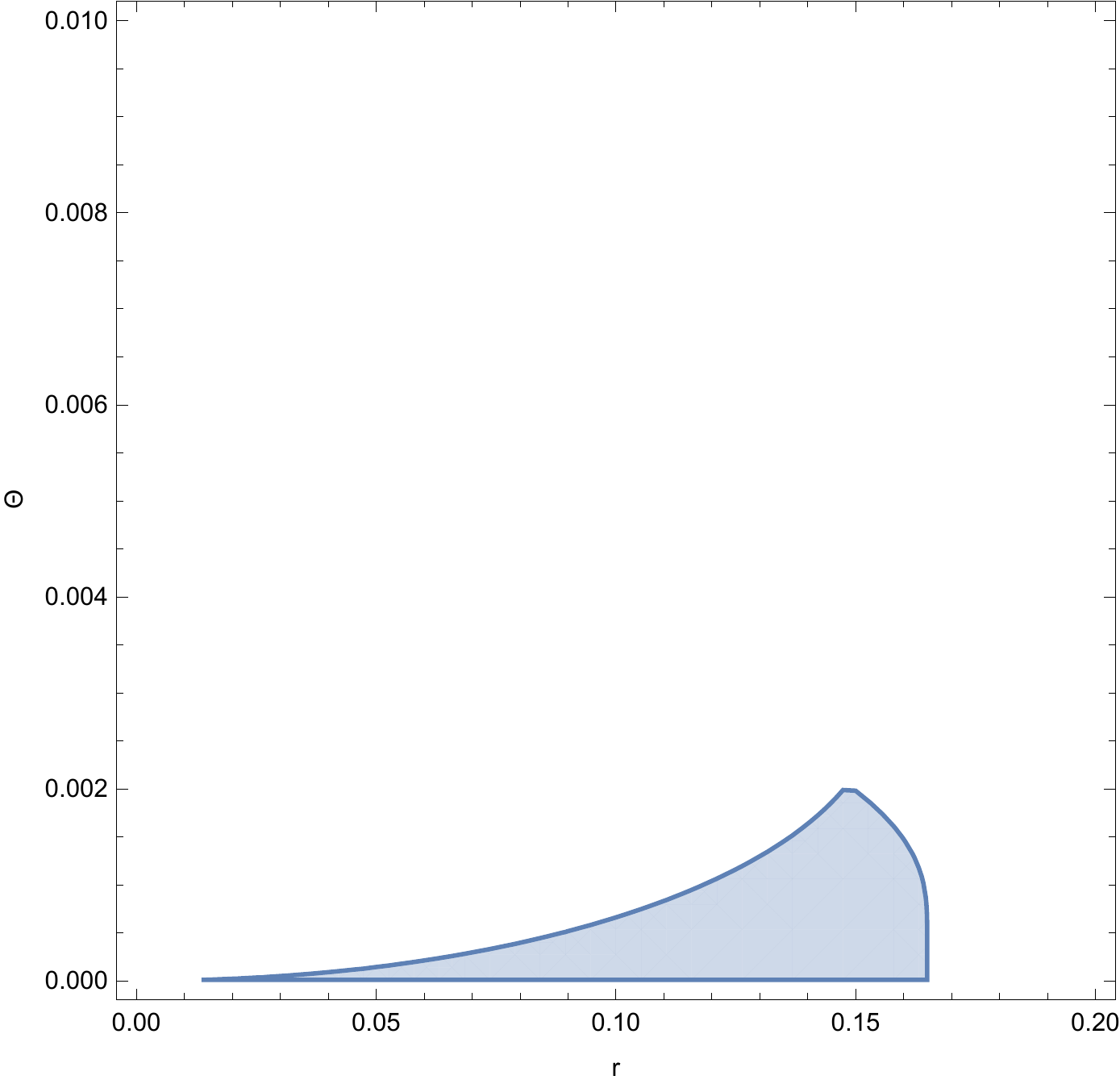}
\includegraphics[width=0.75\columnwidth]{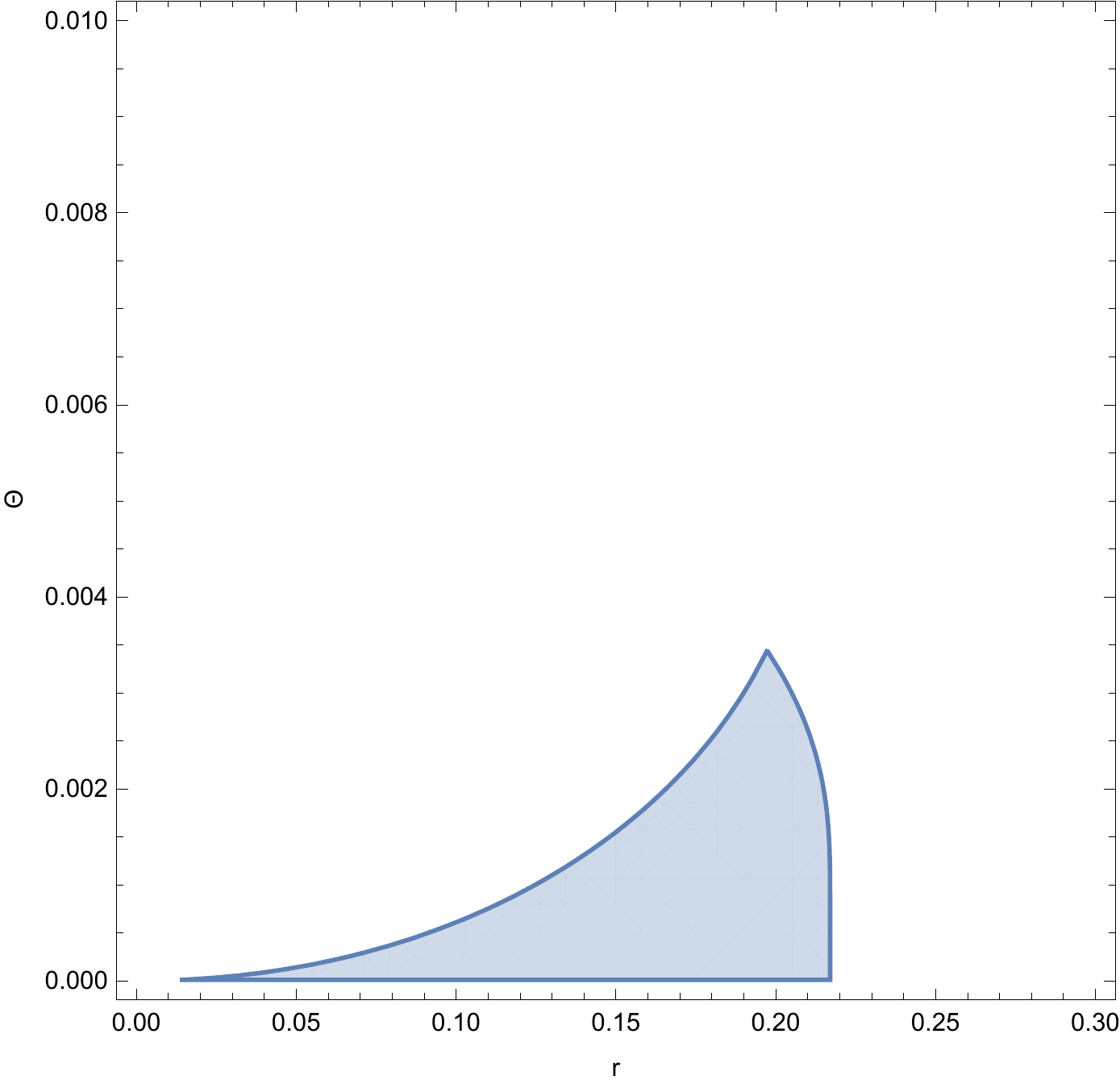}
\caption{Region containing CTC where weak energy condition is preserved for the parameters i) $Q=0.6, a=0.5$ (top) and ii) $Q=0.7, a=0.5$ (bottom).}
\label{fig:ctcweak_reg_1}
\end{figure}

\begin{figure}
\includegraphics[width=0.75\columnwidth]{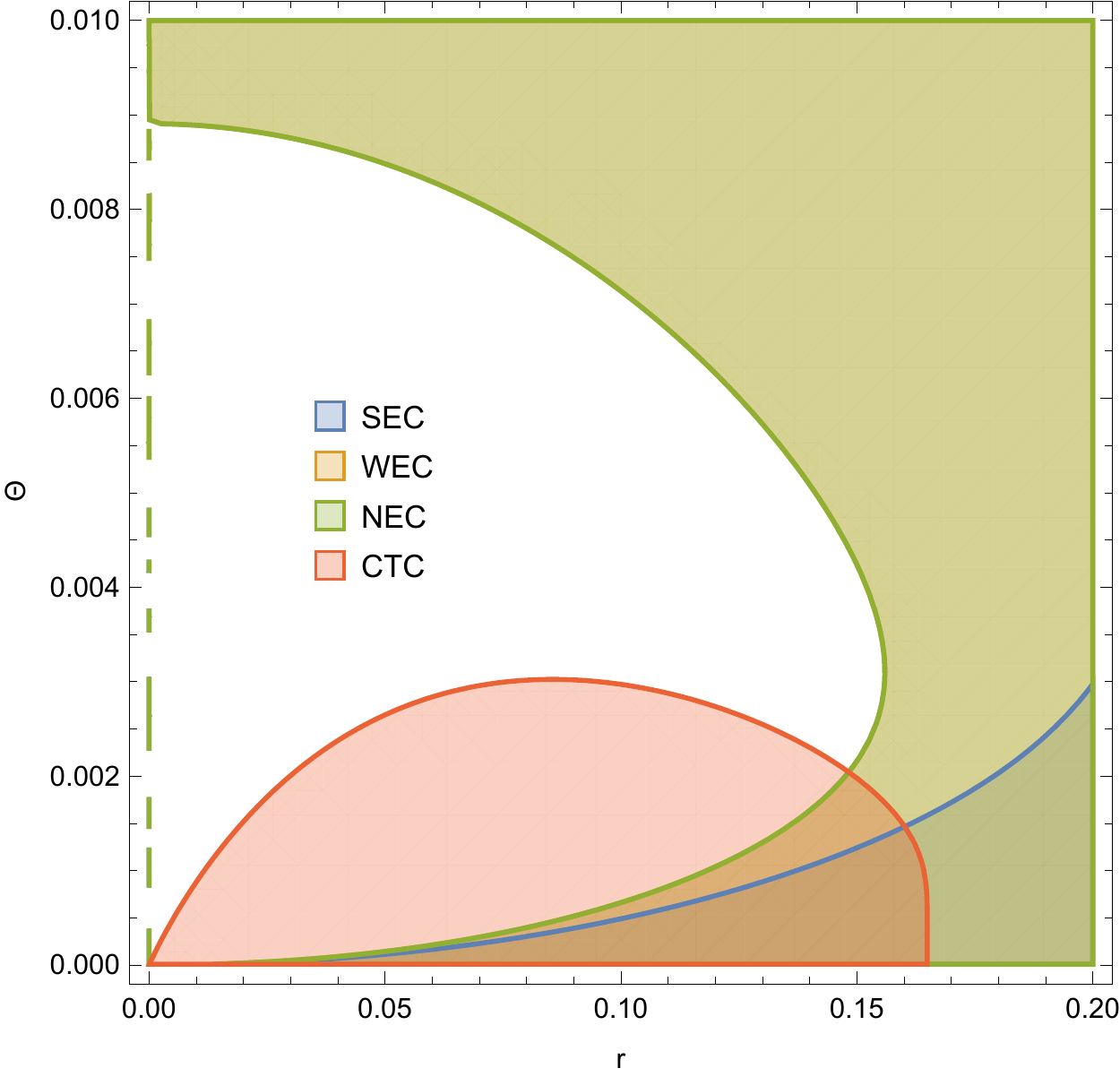}
\includegraphics[width=0.75\columnwidth]{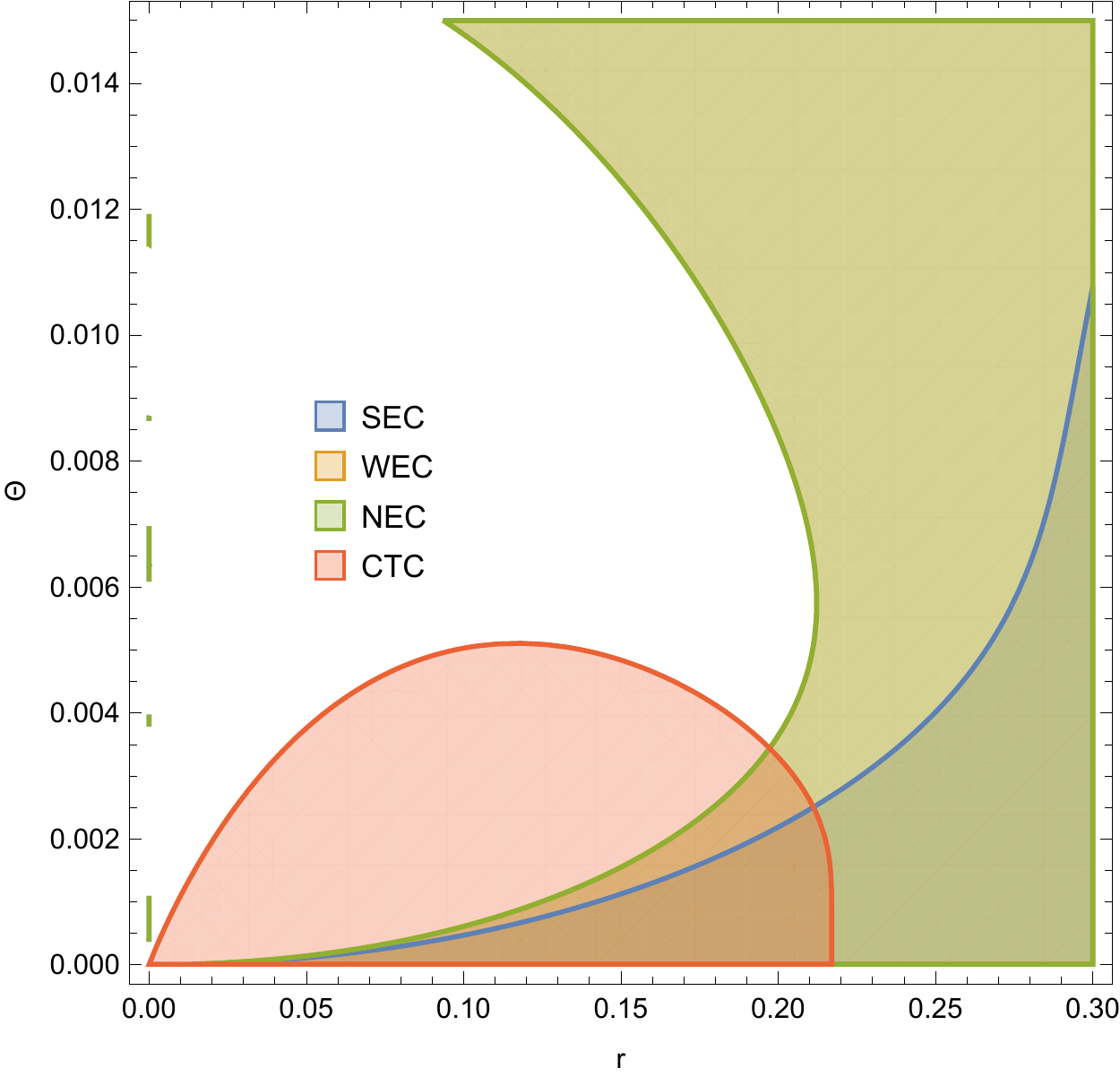}
\caption{Region containing CTC where weak, strong and null energy conditions are preserved for the parameters
i) $Q=0.6, a=0.5$ (top) and ii) $Q=0.7, a=0.5$(bottom).}
\label{fig:CTC-allenergy}
\end{figure}

\begin{figure*}
    \includegraphics[width=0.45\columnwidth]{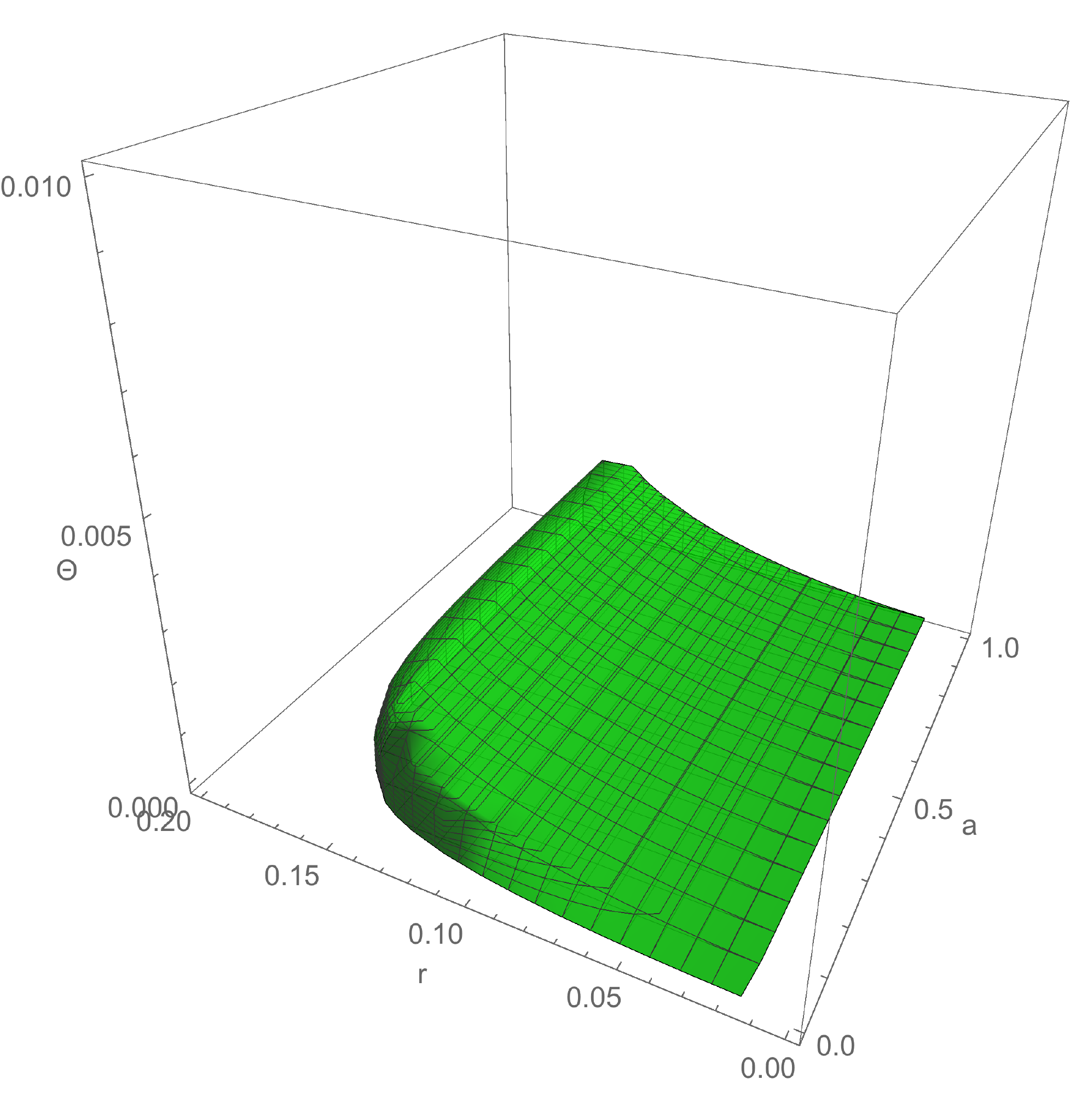}
    \includegraphics[width=0.45\columnwidth]{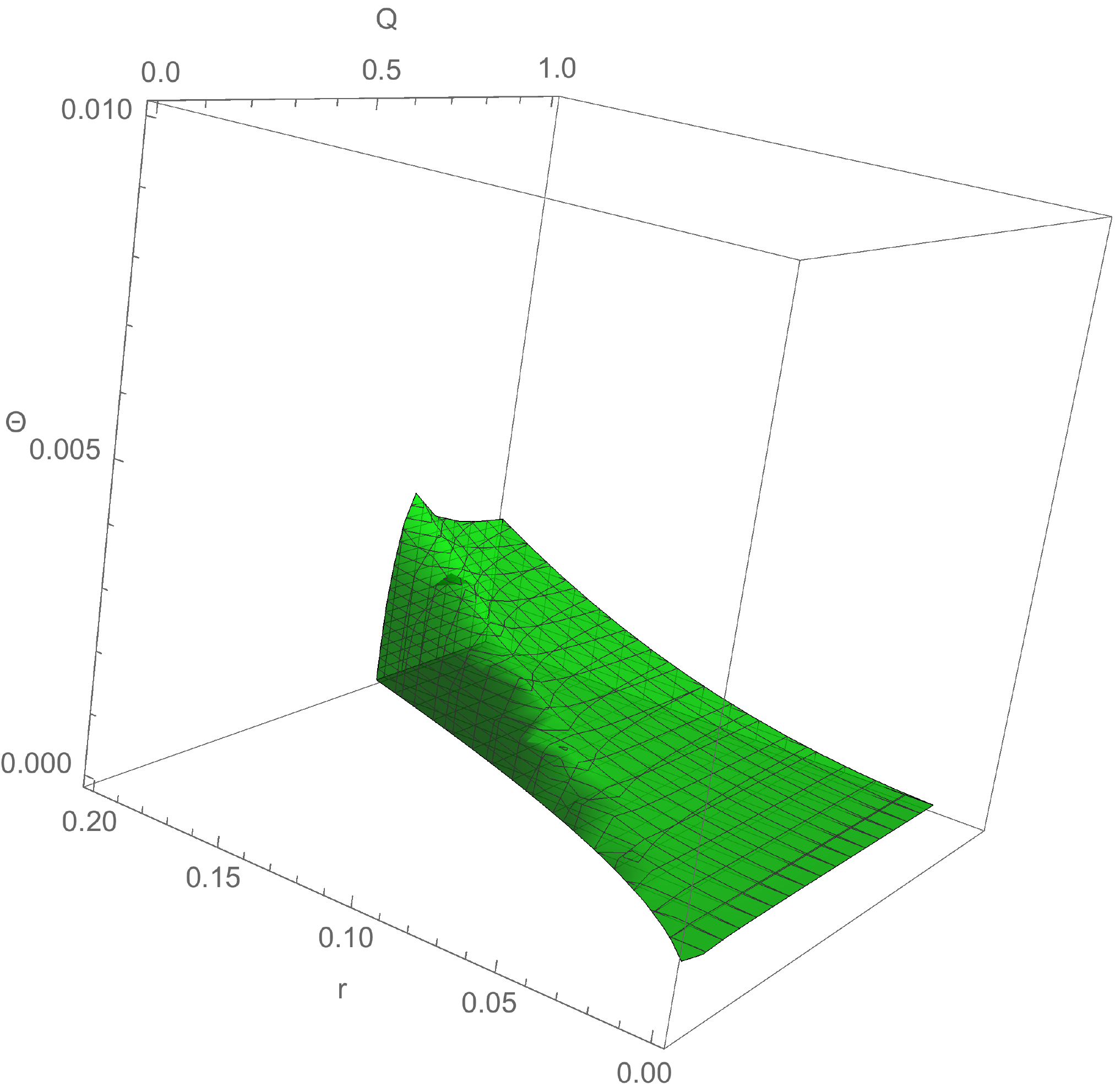}
    \includegraphics[width=0.45\columnwidth]{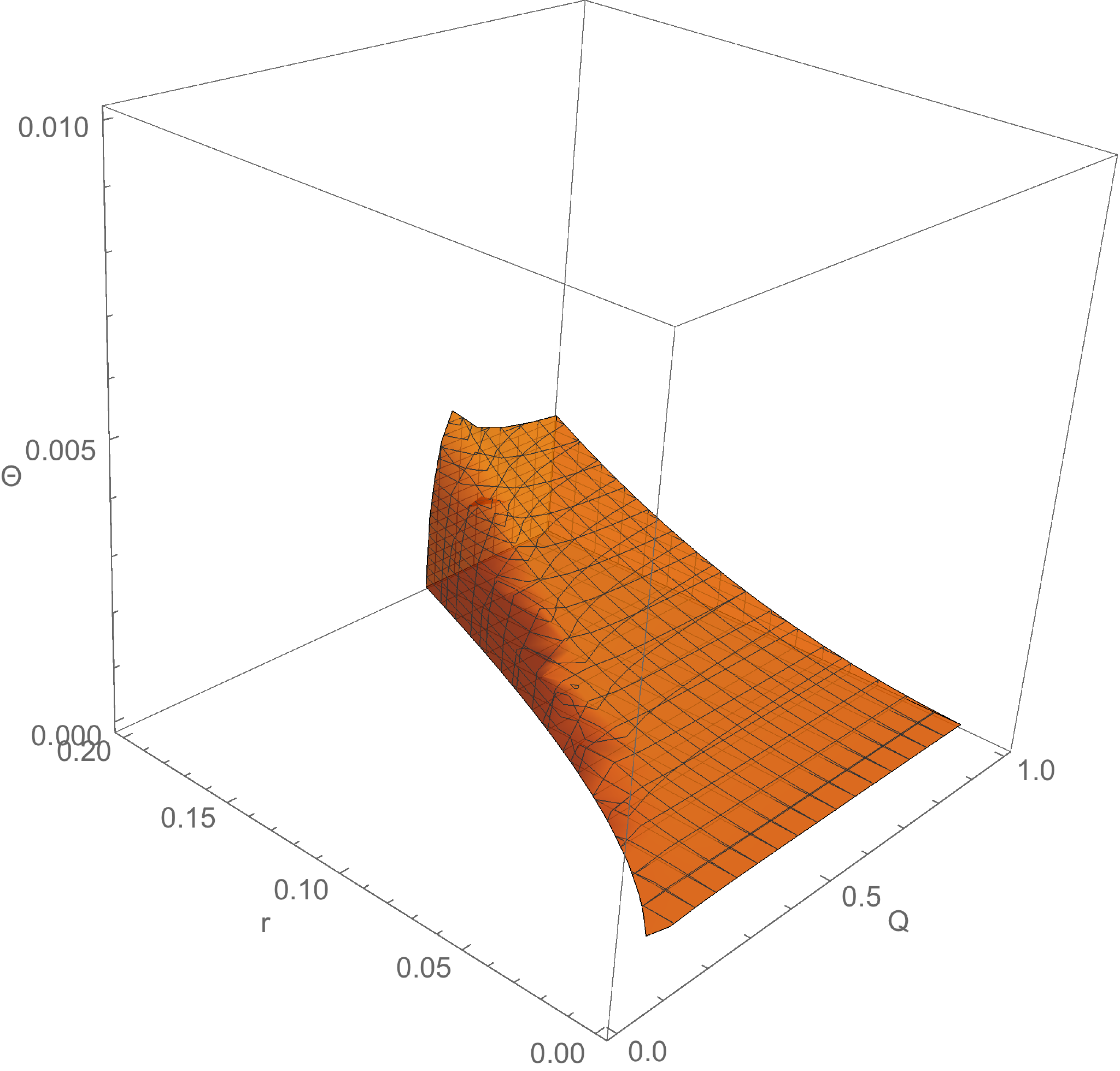}
   \includegraphics[width=0.45\columnwidth]{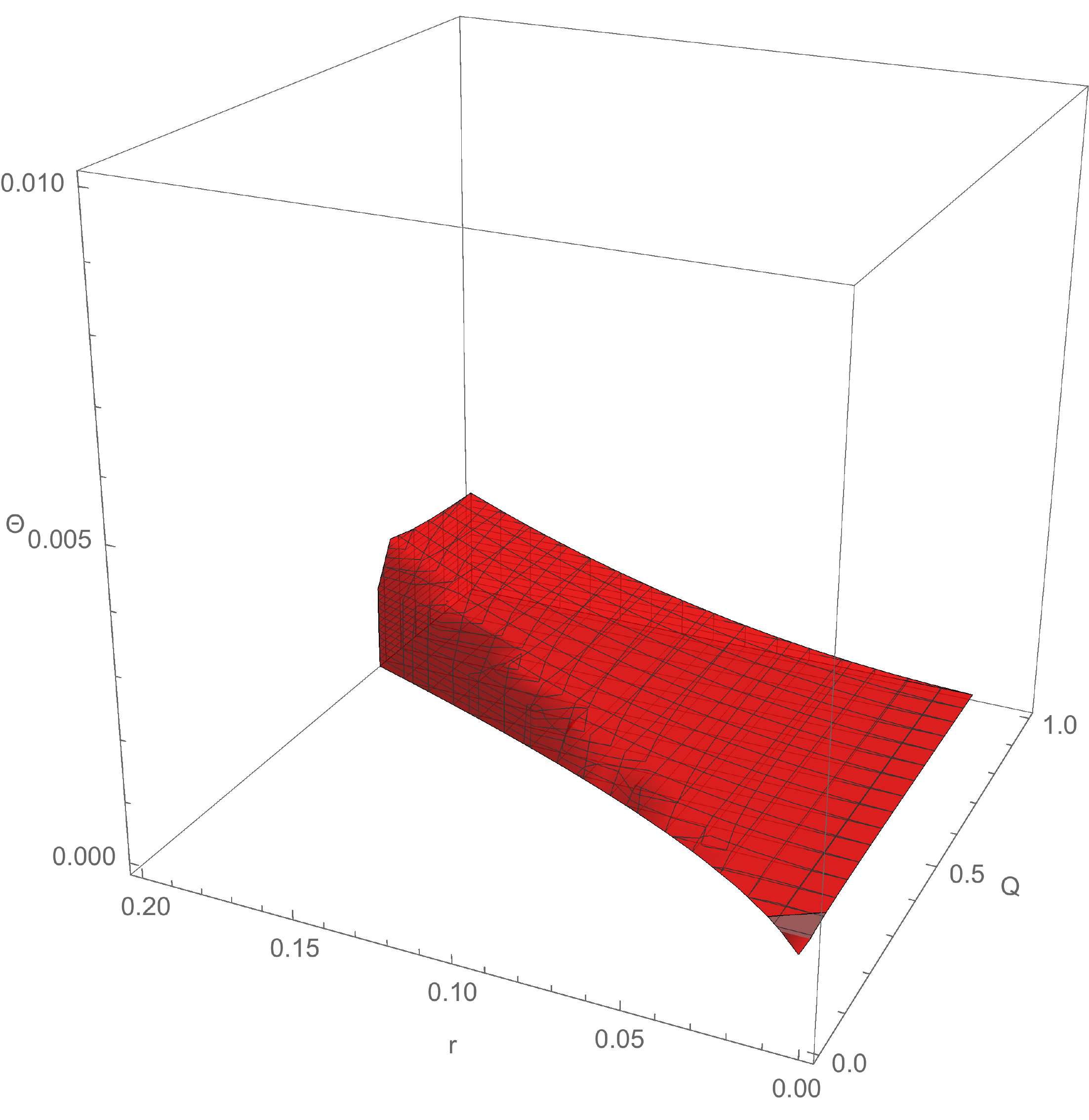}

  \caption{i) $3D-$ region showing the co-existence of CTCs and weak energy condition in $r, \Theta, a$ space with $Q=0.6$ (first plot)  and $r, \Theta, Q$ space with $a=0.5$ (second plot). ii) $3D-$ region showing the co-existence of the CTCs and the null  (third plot) and strong (fourth plot) energy conditions in $r, \Theta, Q$ space with $a=0.5$. }
  \label{fig:weak_r_th_a_null_th_q}
\end{figure*}

We now illustrate the interplay between the weak energy condition and CTCs. Fig.~\ref{fig:ctcweak} illustrates the various quantities , $\rho$, $\rho+p_i$ as a function of radius for the fixed $\Theta, Q$ and $a$. We superimpose the graphs for various expressions that dictate the status of the weak energy condition with the existence of CTCs i.e. $g_{\phi\phi}$. For definiteness we use the parameter set chosen for green curve in Fig.~\ref{fig:ctc_reg_1}. We note that the energy density $\rho$ is negative for small values of radius,  turns positive and eventually becomes zero. Intuitively this is expected since the NCKN is a regular spacetime where singularity is smeared out. We expect a repulsive effect would sustain an equilibrium scenario that has smeared mass in place of a singularity. The other curves are $\rho+p_i$, we see that they become negative closer to the singularity and become positive for a larger radius before eventually becoming zero. It can be seen from the figure that there is an overlap between the radii where $g_{\phi\phi}$ and other weak energy expressions are simultaneously negative.
In Fig.~\ref{fig:ctcweak_reg_1}, we superimpose regions in $r$ and $\Theta$ space where weak energy conditions are preserved and CTCs are present. In the plots, the top one is for parameter values $Q=0.6$ and $a=0.5$ while bottom is  panel has $Q=0.7$ and $a=0.5$. The $x$-axis is the re scaled radius $r/M$ and $y-$ axis represents the non-commutative parameter $\Theta$.  The shaded area in both the plots represents the region where CTCs are present and Weak Energy conditions are preserved. Fig.~\ref{fig:CTC-allenergy} illustrate the common region  where  CTC's exist and various the energy conditions are satisfied in the $r-\Theta$ space (with $Q=0.6, a=0.5$ and $Q=0.7, a=0.5$, respectively). 
 The coloured regions in the parameter space indicate the preservation of the respective energy condition (weak, strong, null). 
Few observations from the plots are in order. We can make a local statement regarding the relation between energy conditions and CTCs. At a given spacetime point, one can have a CTC  passing through it and the weak energy condition at that point may or may not be violated. For a point at radius $r$, we observe that weak energy condition is preserved for small $\Theta$ but get violated as we increase $\Theta$. This trend is also displayed for the elimination of CTCs. This seems to indicate that a violation of weak energy conditions favours the elimination of CTCs. This can be seen from Fig.~\ref{fig:CTC-allenergy}  that the boundary of the CTC region occurs beyond the region of validity of the weak energy condition, strongly suggesting that in the absence of singularities, the violation of weak energy condition help us to eliminate CTCs. The large $r$ behaviour of CTC boundary might not contribute to this correlation since even for KN spacetime, the CTCs die down as radius increases. We now recall Hawking's statement which states that for the spacetimes with no singularities and no pathological behaviour at future infinity, the presence of CTCs requires the violation of weak energy condition. The conclusions drawn from our analysis seems to suggest that the above statement is valid globally but not locally.  It means it might be possible that the weak energy condition could be violated at some region of the spacetime but not necessarily at the same location as that of CTCs. To elucidate this point further, we scan the entire parameter space and look for the regions containing CTCs where weak energy conditions are also preserved. In the first two plots of Fig.~\ref{fig:weak_r_th_a_null_th_q}, we highlight the region in the parameter space:  $r-\Theta-a$ and $r-\Theta-Q$ (right) where the CTCs are present and also the weak energy condition is preserved. For completeness, we have also presented the $3D-$ regions of co-existence of CTCs and null, strong energy conditions in the last two plots of Fig.~\ref{fig:weak_r_th_a_null_th_q}.

These plots indicate that at a given radius, one can make a local statement that for the spacetimes with no singularities and no pathological behaviour at future infinity, the presence of CTCs does not imply the violation of weak energy conditions. Consequently, Hawking's chronological conjecture holds in the global sense but not locally. 
\subsection{Correlation index}
Here we evaluate the correlation index for the NCKN case for fixed point ($t, r, \theta = \pi/2, \phi$) on the equatorial plane in spacetime.
\begin{equation}
\label{cindexnckn}
\mathcal{C}_r=\frac{\bigint \frac{f}{| f|}\left(\frac{g_{\phi\phi}}{|g_{\phi\phi}|}-1\right) da dQ d\Theta
}{\bigint \left(\frac{g_{\phi\phi}}{|g_{\phi\phi}|}-1\right) da dQ d\Theta}
\end{equation}
with the function $f$ given as  $$f = Min[\rho, \rho+p_1, \rho+p_2, \rho+p_3]$$. 
\begin{figure}
  \includegraphics[width=1.0 \columnwidth]{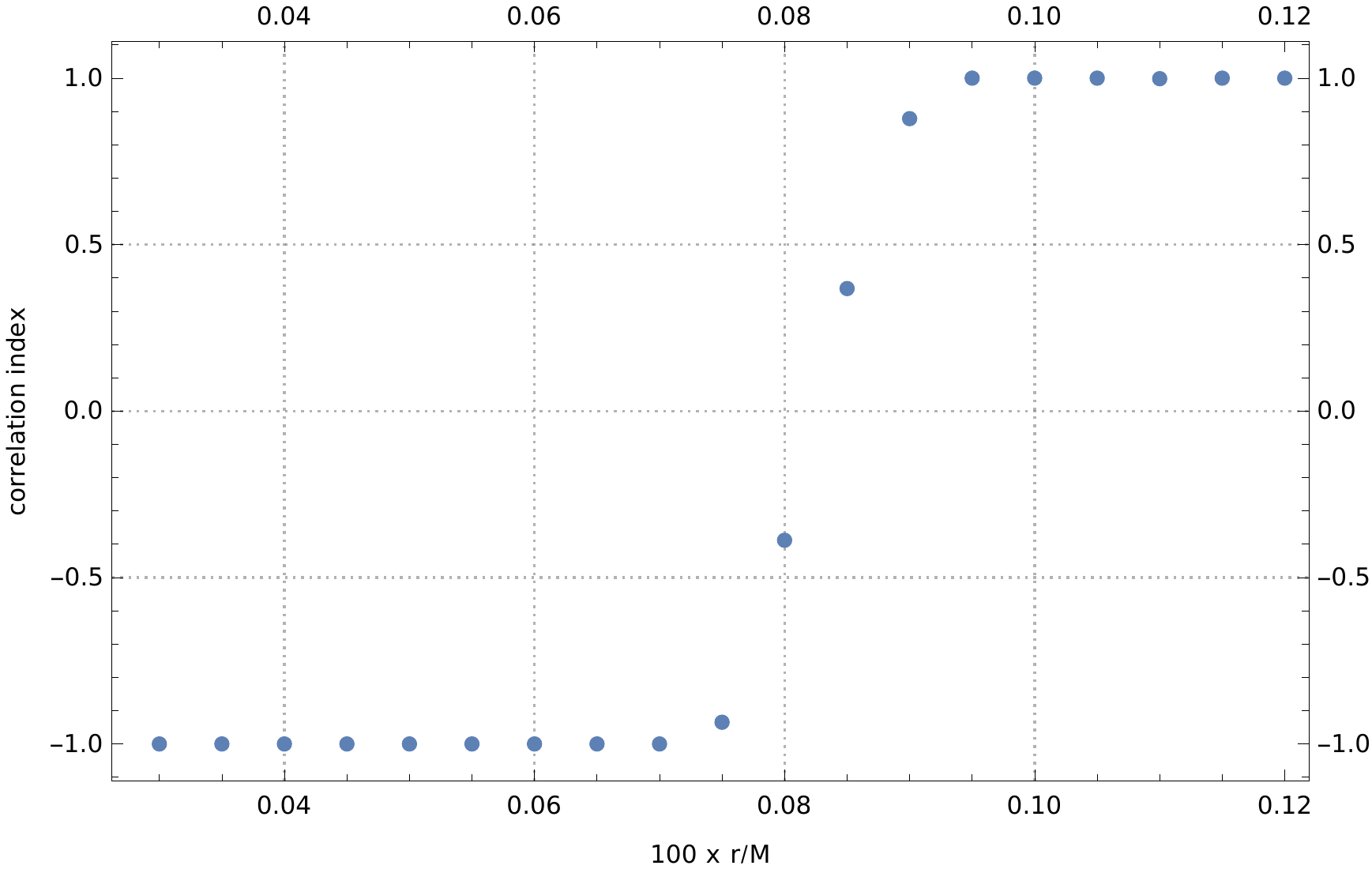}
  \caption{Indices for correlation between WEC and CTCs in the NCKN geometry for values of x ranging $0.03$ to $0.200$ (scaled by factor $10^{2}$). The index is calculated over $a$, $Q$ and $\Theta$.}
  \label{fig:nc1index}
\end{figure}

In the Fig.~\ref{fig:nc1index}, we present the correlation index at few values of $x=r/M$. The index is calculated by integrating over the parameter space involving $a,Q, \Theta$. The range of $x$ chosen is such that the CTCs are present in some parameter range. The normalized radius $(x=r/M)$ is varied from $0.03$ to $0.12$. The index is evaluated by carrying out an integration over $a$, $Q$ and $\Theta$. We see that for smaller values of x, the index is close to $-1$ and for relatively larger values of x, the index smoothly varies till it becomes $+1$ for larger values of $x$. This implies that for smaller values of x, the CTCs exist in an environment where WEC is violated. The effect is more prominent in smaller values of x because the WEC violation is required to smoothen out the singularity. The NCKN spacetime asymptotes to the standard Kerr-Newman spacetime for larger $r$ where the energy conditions are not broken. So for larger x, the index smoothly goes to the WEC energy preserving environment.   \\

\section{Rastall gravity}\label{sec:rastall}
We now consider the interesting Kerr-Newman solution in Rastall gravity. This model presents a contrasting situation where singularity is present and parameter range can be chosen where CTCs are present and another parameter range where they are eliminated. Rastall gravity is set up based on the following framework. The conservation of energy-momentum tensor plays an important role in classical general relativity. In Rastall gravity, there is a relaxed assumption of conservation condition of the energy-momentum tensor \cite{Rastall}. Here the divergence of energy-momentum tensor is given as,
\begin{align}
\nabla_{\nu}T^{\mu\nu}=\lambda \nabla^\mu R
\end{align}
Where R is the Ricci scalar. We see that in the flat spacetime where the Ricci scalar vanishes, one recovers the usual conservation law. $\lambda$ is called the Rastall parameter and measures a deviation from the standard classical general relativity's conservation law. The field equation in Rastall gravity is given by,
\begin{align}
G_{\mu\nu}+\kappa\lambda g_{\mu\nu}=\kappa T_{\mu\nu}
\end{align}
When $\lambda\rightarrow 0$,  we get $\kappa=8\pi G_{N}$  which is the Newtonian coupling constant while for non zero value of $\lambda$, it is called Rastall coupling constant. By solving the modified field equation we can get the metric for different types of spacetimes.

\subsection{Kerr-Newman spacetime in Rastall gravity}
The complete derivation of the Reisnner-Nordstrom spacetime in Rastall gravity surrounded by perfect fluid have been studied in \cite{Darabi}. 
The metric for Kerr-Newman spacetime in Rastall gravity surrounded by a perfect fluid can be given by using the 
\begin{widetext}
\begin{eqnarray}
ds^2 &=& -\left(1-\frac{2Mr}{\rho^2}+\frac{Q^2}{\rho^2}-\frac{N_s r^{\frac{1+3\omega_s-6\kappa\lambda(1+\omega_s)}{1-3\kappa\lambda(1+\omega_s)}}}{\rho^2}\right)dt^2
-2a\sin^2\theta\left(\frac{2Mr-Q^2+N_s r^{\frac{1+3\omega_s-6\kappa\lambda(1+\omega_s)}{1-3\kappa\lambda(1+\omega_s)}}}{\rho^2}\right)dtd\phi
\nonumber\\
&+& \frac{\rho^2}{\rho^2-2Mr+Q^2-N_sr^{\frac{1+3\omega_s-6\kappa\lambda(1+\omega_s)}{1-3\kappa\lambda(1+\omega_s)}}+a^2sin^2\theta}dr^2
+\rho^2d\theta^2 \nonumber\\
&+& sin^2\theta\left[r^2+a^2
+ a^2sin^2\theta\left(\frac{2Mr-Q^2+N_s r^{\frac{1-3\omega_s}{1-3\kappa\lambda(1+\omega_s)}}}{\rho^2}\right)\right]d\phi^2
\label{rastallkn}
\end{eqnarray}
\end{widetext}
where $\omega_s$ is the state parameter of the fluid surrounding the black hole and $N_s$ represents the surrounding field structure parameter. Therefore the Horizon function from Eq.~\eqref{rastallkn} is given as 
\begin{align}
\Delta(r) =\rho^2-2Mr+Q^2-N_s r^{\frac{1+3\omega_s-6\kappa\lambda(1+\omega_s)}{1-3\kappa\lambda(1+\omega_s)}}+a^2sin^2\theta
\end{align}
Similar derivation can also be found in \cite{Xu} where the Kerr-Newman spacetime is derived in Anti-De Sitter spacetime. To find the CTC we require the $g_{\phi\phi}$ term. From eqn(13), the $g_{\phi\phi}$ in Rastall gravity surrounded by a perfect fluid is given as

\begin{figure*}
    \includegraphics[width=0.68\columnwidth]{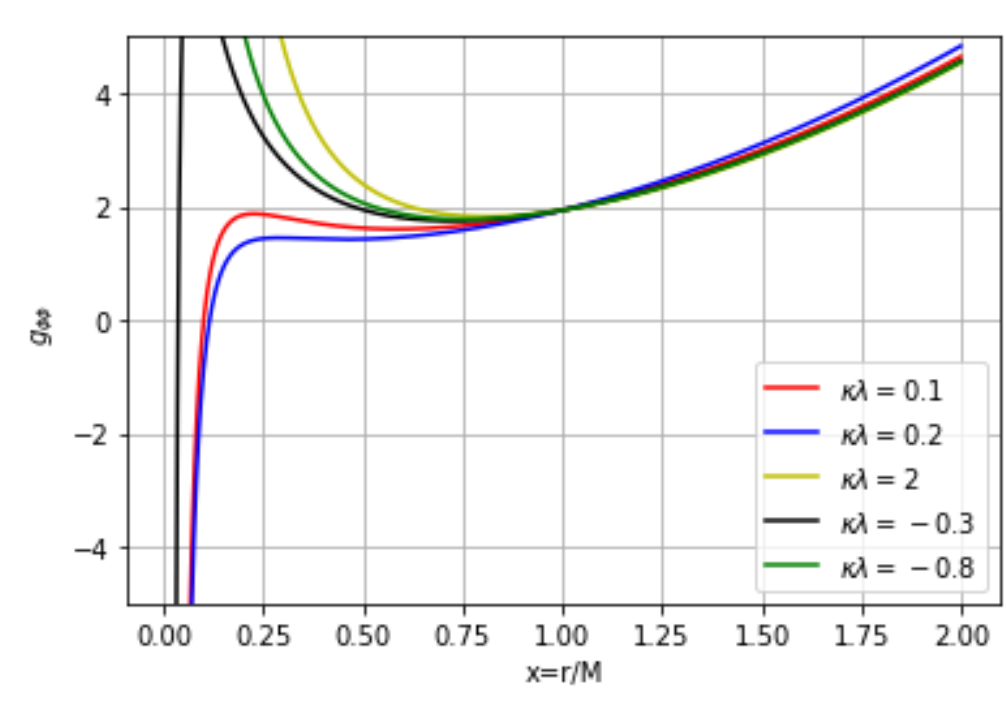}
    \includegraphics[width=0.68\columnwidth]{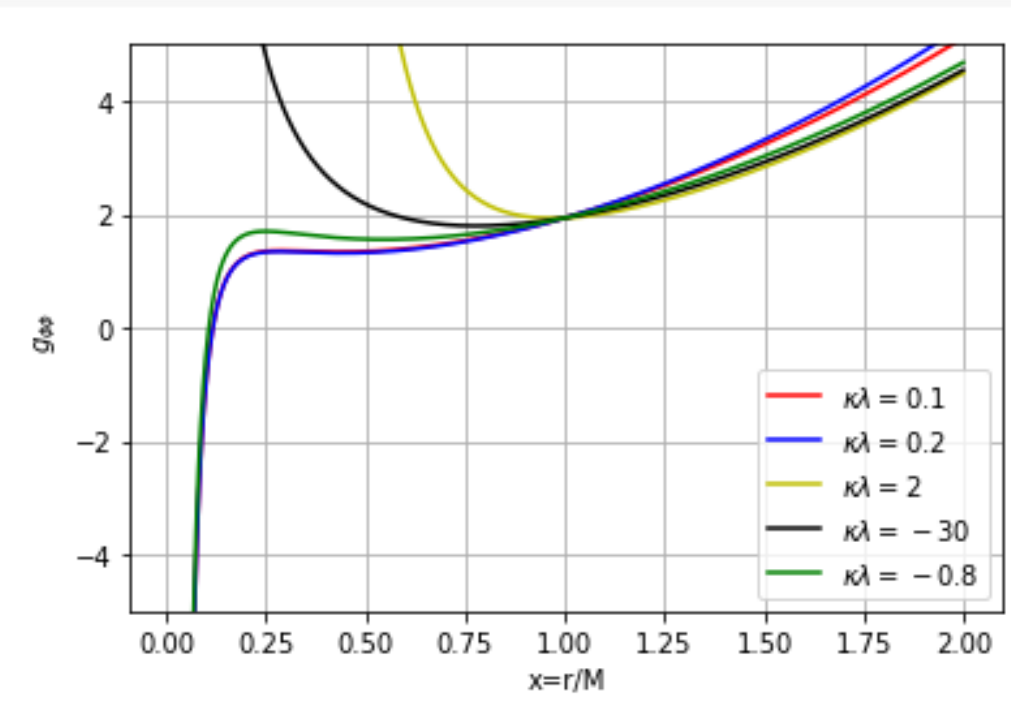}
    \includegraphics[width=0.68\columnwidth]{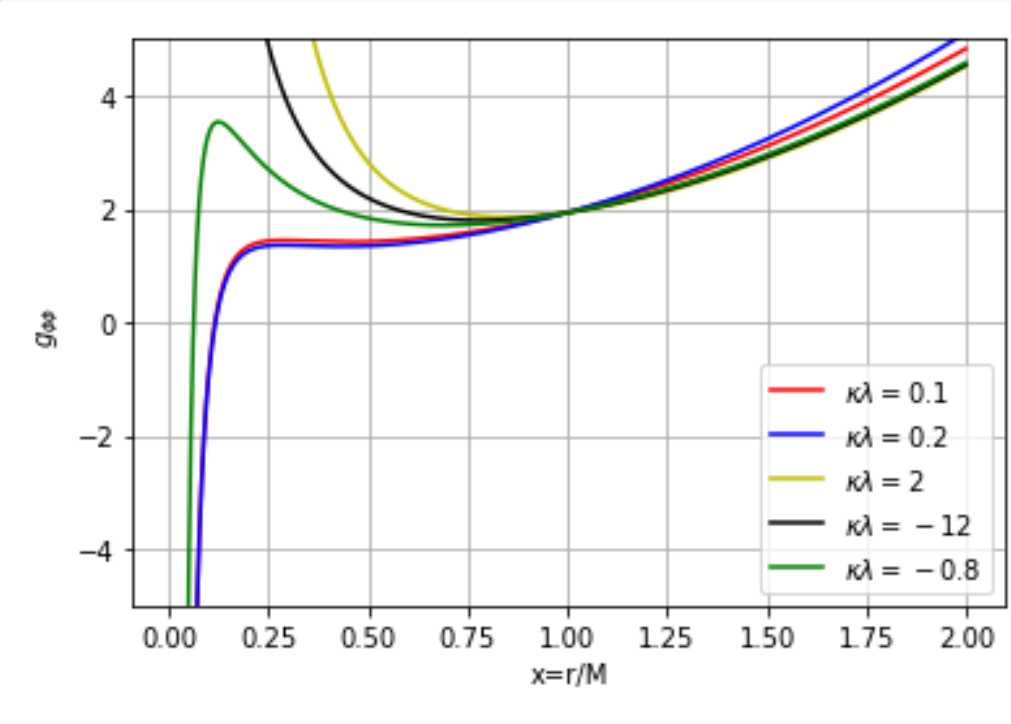}
  \caption{ CTC plot on the equatorial plane for the parameters: i) $a=0.5M$, $q=0.25M$, $\omega_s=0$ and $N_s=1$  ii)  $a=0.5M$, $q=0.25M$, $\omega_s=-\frac{2}{3}$ and $N_s=1$ and iii)  $a=0.5M$, $q=0.25M$, $\omega_s=-\frac{1}{3}$ and $N_s=1$.}
  \label{fig:figurerastallomega}
\end{figure*}

\begin{widetext}
\begin{eqnarray}
g_{\phi\phi}= sin^2\theta\left[r^2+a^2+a^2sin^2\theta \left(\frac{2Mr-Q^2+N_s r^{\frac{1-3\omega_s}{1-3\kappa\lambda(1+\omega_s)}}}{\rho^2}\right)\right] \ ; \ \rho^2 &=& r^2 + a^2 cos^2\theta
\label{rastallgphiphi}
\end{eqnarray}
\end{widetext}
%
We see that there are two independent parameters that can be examined in the context of CTCs ($\kappa\lambda$ and $\omega$). We now look for CTC in Kerr-Newman black hole for various equations of state.  The value of $\omega_s$ characterizes each fluid  (for e.g dust ($\omega_s=0$), dark energy ($\omega_s=-\frac{2}{3}$) and perfect dark matter ($\omega_s=-\frac{1}{3}$)). We substitute this value into Eq.~\eqref{rastallgphiphi} and plot it in the equatorial plane against $x=r/M$ as shown in Fig.~\ref{fig:figurerastallomega} for various values of $\kappa \lambda$. We see that for each value of $\omega$, we can find values of $\kappa\lambda$ for which CTCs are there since $g_{\phi\phi}$ becomes negative and also values of $\kappa\lambda$ for which $g_{\phi\phi}$ stays positive. We can therefore see that the model is equipped with a natural protection mechanism against causality violation.
\par
\begin{figure*}
    \includegraphics[width=0.68\columnwidth]{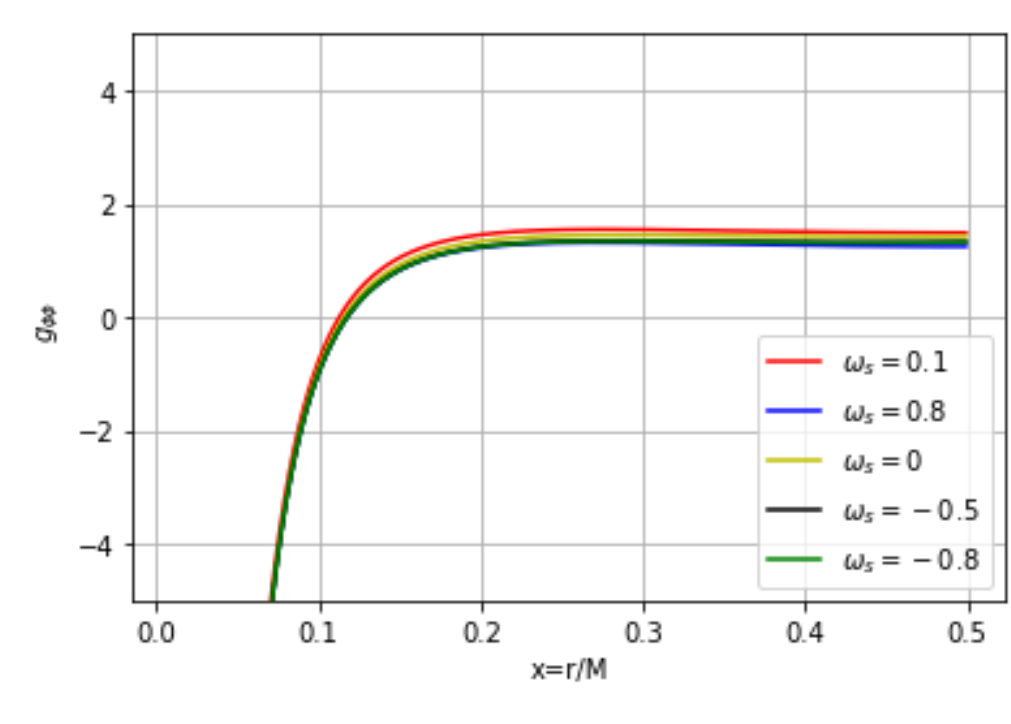}
    \includegraphics[width=0.68\columnwidth]{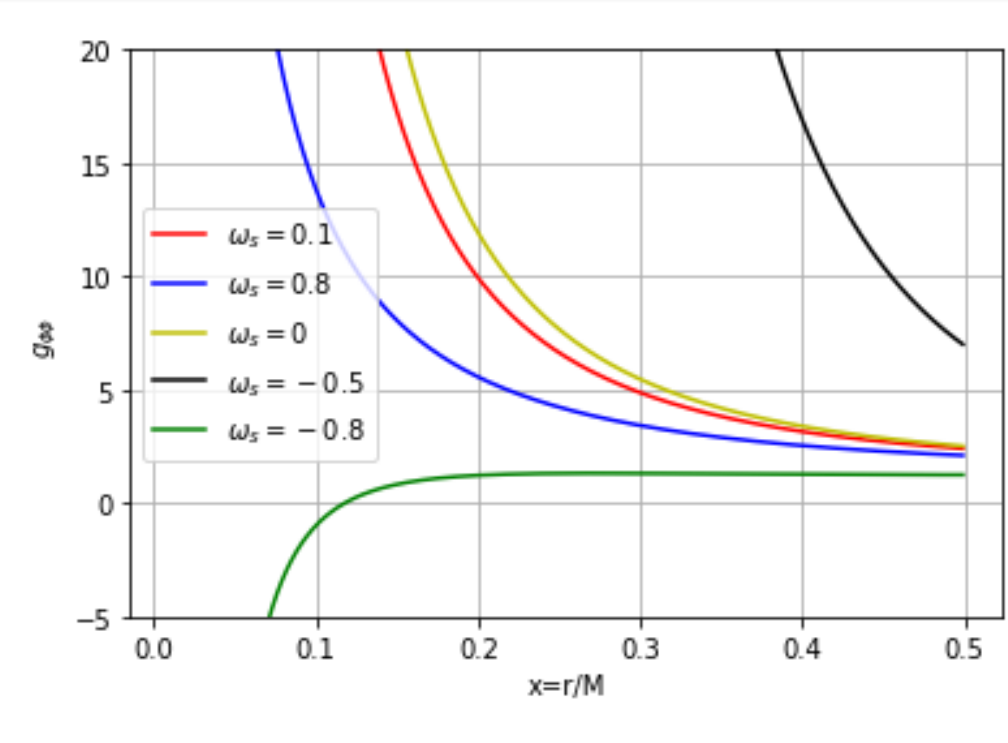}
    \includegraphics[width=0.68\columnwidth]{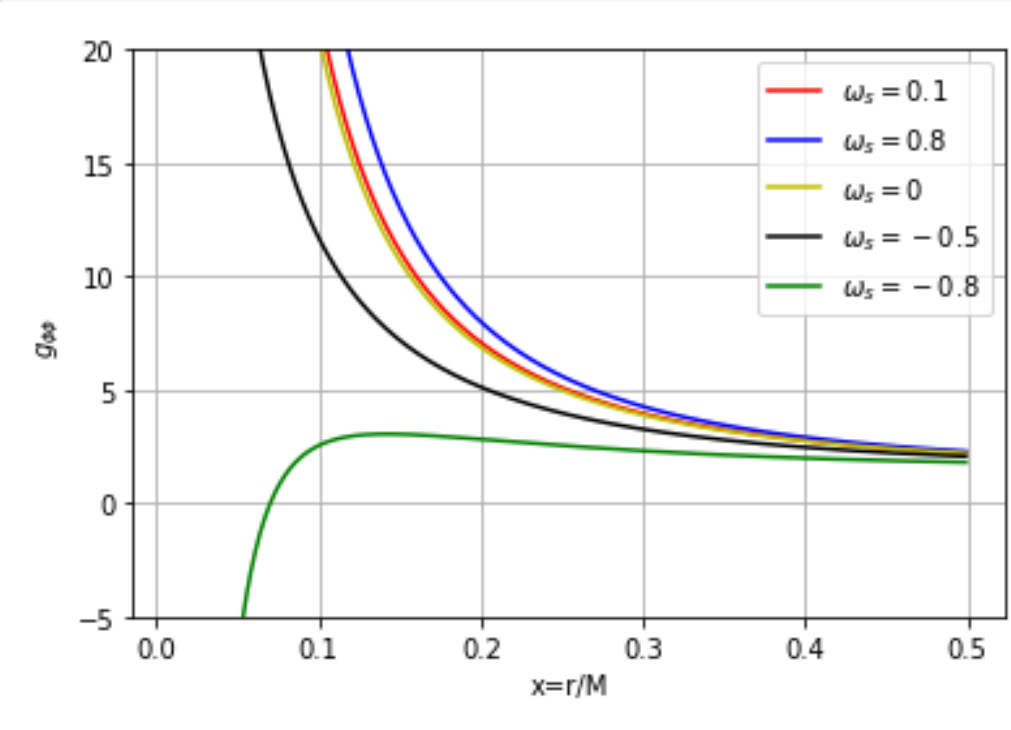}
  \caption{ CTC plot on the equatorial plane for the parameters: i) $a=0.5M$, $q=0.25M$, $\kappa\lambda=0.2$ and $N_s=1$  ii)  $a=0.5M$, $q=0.25M$, $\kappa\lambda=1.33$ and $N_s=1$ and iii)  $a=0.5M$, $q=0.25M$, $\kappa\lambda=-5$ and $N_s=1$}
  \label{fig:figurerastallkappa}
\end{figure*}

In the second part , Fig.~\ref{fig:figurerastallkappa},  we  keep $\kappa\lambda$ constant and plot $g_{\phi\phi}$ component for different values of $\omega_s$. As we can see from Fig.~\ref{fig:figurerastallkappa},  for $\kappa\lambda=0.2$ there are CTC for every value of $\omega_s$ portrayed in the figure, but when we increase the value of $\kappa\lambda$ to 1.33 then CTC is seen only for $\omega_s=-0.8$. This implies that if we go on increasing the value of $\kappa\lambda$ there is a value above which for a given $\omega_s$ value there shouldn't be any CTC. This behaviour is not seen for $\omega_s=-1$ for which how much bigger the value of $\kappa\lambda$ is there is always a CTC. This similar behaviour is also seen for the negative value of $\kappa\lambda$. We see that while varying $\omega$, we can have the possibility of eliminating CTCs. We, therefore, have two handles that can be tweaked to choose scenarios that aid in eliminating CTCs. 

\subsection{Correlation index of CTCs with weak energy condition}
\begin{figure}
  \includegraphics[width=1.0 \columnwidth]{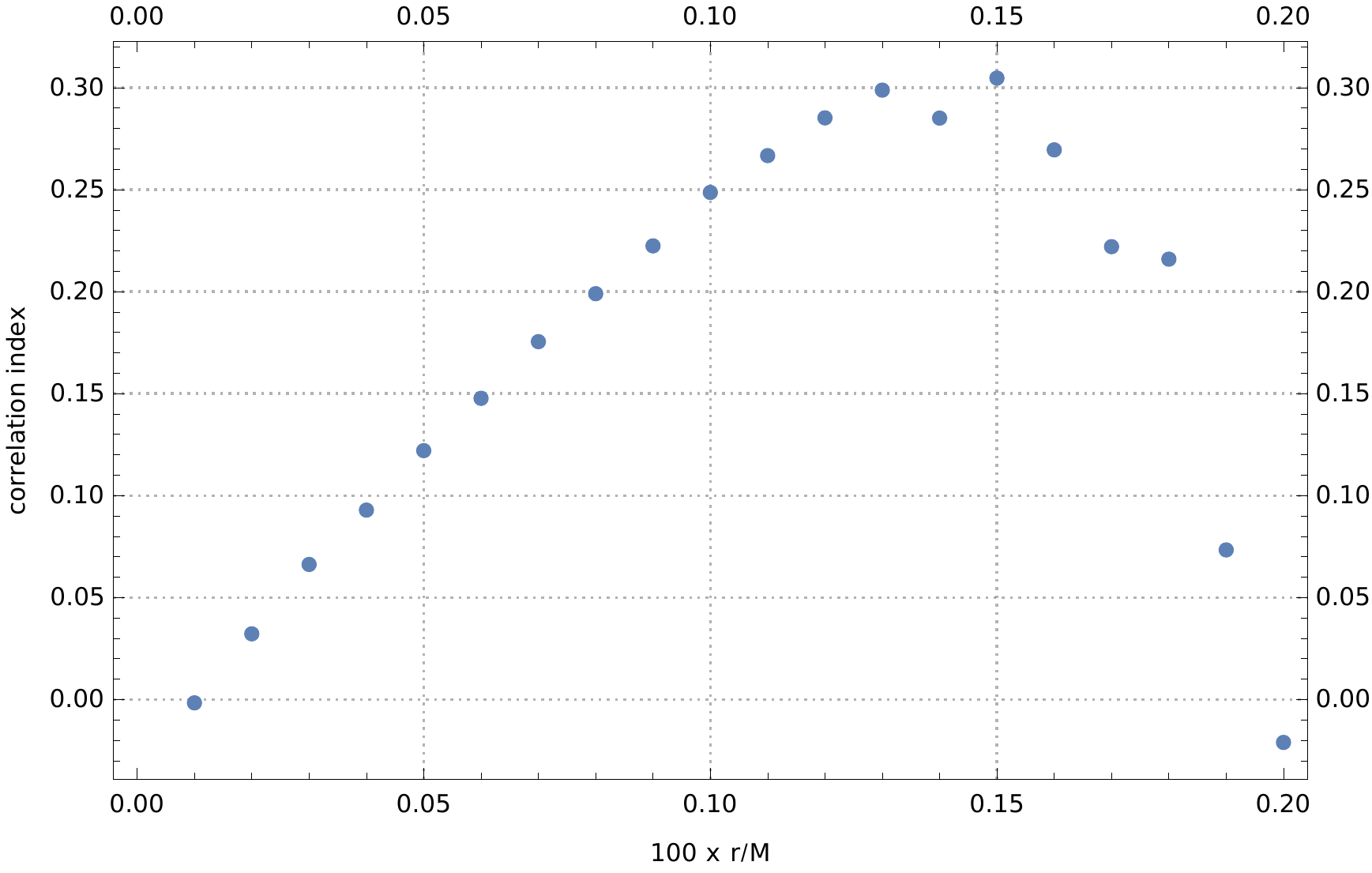}
  \caption{Indices for correlation between WEC and the existence of CTCs in the Kerr-Newman spacetime of Rastall gravity for values of x (scaled by factor $10^{2}$) from $0.01$ to $0.2$.}
  \label{fig:xindex}
\end{figure}

The expressions for the weak energy condition in the model under consideration is given in \cite{Darabi,Xu}. The condition has a simple form given by $(3\kappa\lambda(1+\omega)-3\omega)(1-4\kappa\lambda)\geq 0$. If the above inequality is met, it implies WEC is preserved and WEC is broken in the parameter space where the inequality is not met. We now adapt the index we had defined in Eq.~\eqref{cindex2} to the case of Rastall gravity. Firstly we fix $\theta=\pi/2$ and $\phi=0$. Then for a fixed value of radius (normalized with the mass parameter M) $x=r/M$, we scan the entire ecosystem (parameter space) consisting of angular momentum $a$, the charge parameter $Q$, equation of state parameter $\omega$ and Rastall parameter $\kappa$. We then define the index as,
\begin{equation}
\label{cindexrstall}
    \mathcal{C}_r=\frac{\bigint \frac{f}{| f|}\left(\frac{g_{\phi\phi}}{|g_{\phi\phi}|}-1\right) da dQ d\omega d\kappa
    }{\bigint \left(\frac{g_{\phi\phi}}{|g_{\phi\phi}|}-1\right) da dQ d\omega d\kappa}
\end{equation}
The index $r$ indicates that this calculation is at a fixed radial coordinate $r$. $f$ is the function whose sign decides whether the weak energy condition is broken or not. We note that when we freeze $r$, it does not necessarily imply that we are performing the integral over a fixed spacetime point. The reason is that the spacetime manifolds are different when we choose a different set of parameters, and there is little meaning in identifying spacetime points across different spacetimes. So in a sense, we are averaging over different spacetime solutions. Given below is the value of $\mathcal{C}_r$ over few values of $r$ that definitely have CTCs in some parameter range. Otherwise the integral becomes undefined since both denominator and numerator go to $zero$, in which case numerical analysis yields spurious values. The observations are the following. The index whose allowed values are in the interval $[-1,1]$ is seen to be positive in Fig.~\ref{fig:xindex}. 
\begin{figure}
  \includegraphics[width=1.0 \columnwidth]{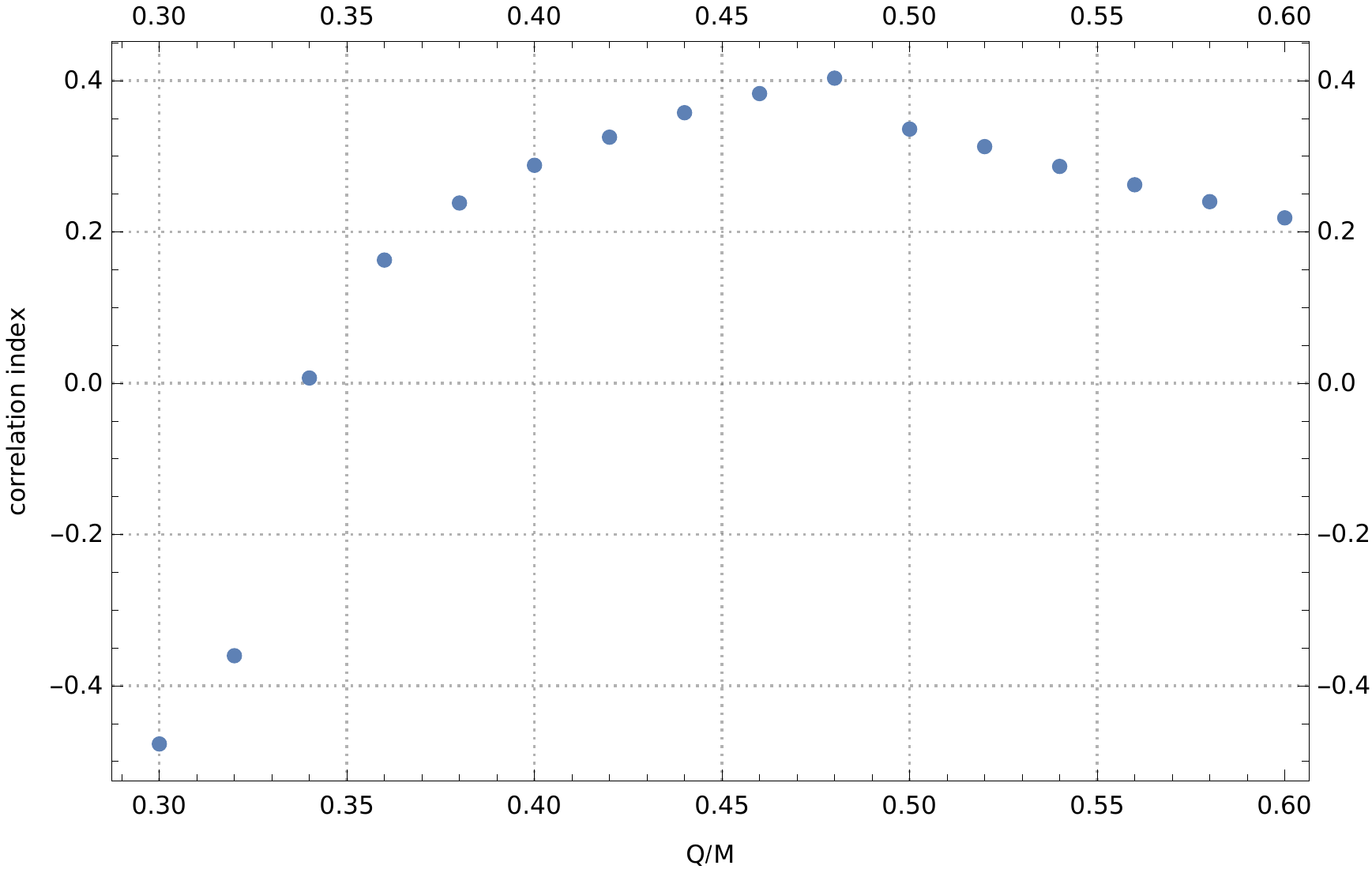}
  \caption{Indices for correlation between WEC and the existence of CTCs in the Kerr-Newman spacetime of Rastall gravity a fixed value of $x=0.15$ and Q goes from $0.3$ to $0.6$ in steps of 0.02}
  \label{fig:qtable}
\end{figure}

This strongly suggests that in the vicinity of a curvature singularity, the CTCs tend to flourish when the WEC is preserved and if the goal is to eliminate CTCs, the breaking of WEC is effective. 
In Fig.~\ref{fig:qtable}, we fix the value of $r$ ($x=0.15$, where the CTCs are definitely present for a large volume in the parameter space). We tabulate the value of the index for few values of $Q$. The above index Eq.~\ref{cindexrstall} is modified such that the integral is over $a$, $\omega$ and $\kappa$ and not Q. We may define this index as $C_{rQ}$ with the subscript indicating that we have evaluated the index for a fixed r and Q. In Fig.~\ref{fig:qtable}, it is seen that the index is positive, indicating again that in the vicinity of a singularity, a violation of WEC helps in the elimination of CTCs.

\section{Conclusions}\label{sec:con}
It was conjectured by Hawking that nature prevents the appearance of CTCs. The problem due to the presence of CTCs is a fundamental pathology in a manner similar to the pathology of singularity. A concrete proof to the question of whether nature prevents CTCs might have to wait for inputs from Quantum Gravity in a manner similar to the cosmic censorship hypothesis. We in this work address few aspects that add to our understanding regarding CTCs. The guiding principle was Hawking's statement regarding the relation between CTCs, singularities and energy conditions. In particular, we study the local validity of Hawking's statement regarding CTCs and energy conditions by defining and computing a correlation index.  We analyze two models that provide contrasting situations. The non-commutative Kerr-Newman black hole is non-singular whereas the Rastall gravity is singular. These two models provide an excellent platform to understand the dependence between CTCs and energy conditions. To quantify the correlations, we define a correlation index between CTCs and energy conditions that averages over all the parameters of the model in the region containing CTCs

Based on the analysis, we could demonstrate that Hawking's statement can have a local and global version. We could show that Hawking's statement can be true globally but need not hold locally. This is manifest in the cases analyzed in this work on both non-commutative and Rastall versions of Kerr-Newman black holes. The analysis of the non-commutative Kerr-Newman black hole yields that in the absence of a singularity, the weak energy condition may or may not be violated at the same spacetime point that has a CTC passing through it. The violation of WEC is conducive in smearing out the singularity. We observe that, based on the correlation index, in the absence of a singularity, the CTCs thrive in a WEC violating environment close to the smeared out singularity. For CTCs that are away from the smeared out singularity, the thriving environment for the CTC is a WEC preserving situation. A violation of WEC favours elimination of CTCs in these regions. An analysis of Rastall gravity presents a contrasting situation where a curvature singularity is in causal contact with the spacetime points having CTCs. Based on calculating the correlation index, it is clear that the thriving environment for the CTCs is a WEC preserving environment. This analysis is also suggestive that violation of energy conditions aid CTC elimination. In effect, this is contrary to the local version of Hawking's statement wherein the WEC is violated in spacetimes that has CTCs but has no singularities or pathological behaviour at infinity.\\

One can explore other models of Kerr-Newman spacetimes (the study in progress)  and other spacetimes admitting the CTCs to gain further understanding of the interplay between CTCs and weak energy condition.
 The study can provide insights and directions towards a more systematic and rigorous study for characterizing  CTCs. One can expect these studies to eventually lead to developing theorems for CTCs in a manner similar to the singularity theorems.
Another area that shall be worth studying is how to extrapolate the aforementioned correlation at the semi-classical level. This might be achieved by 
replacing $T^{\mu\nu}$ with its expectation value corresponding to appropriately chosen quantum states. We would like to address these issues in near future.   
\vspace{-0.2cm}
\section*{Acknowledgements:}
\vspace{-0.5cm}
S.K. is supported by University Grants Commission's Faculty Recharge Programme (UGC-FRP), Govt. of India,  New Delhi, India. V.P. is supported by Shyama Prasad Mukherjee Fellowship, CSIR, India. Some computations were carried out using the resources of IUCAA, Pune, India. 
\par
All the authors contributed equally to this work and the authors' names are mentioned according to the alphabetical order of their last names.

\end{document}